\documentstyle[11pt,aaspp4]{article}
\def\msun{M_{\odot}}
\def\Lsun{L_{\odot}}
\def\rsun{R_{\odot}}
\begin{document}
\title{The Internal Structural Adjustment due to Tidal Heating of
Short-Period Inflated Giant Planets}
\author{Pin-Gao Gu}
\affil{Institute of Astronomy and Astrophysics, Academia Sinica,
Taipei 106, Taiwan, R.O.C.}
\and
\author{Peter H. Bodenheimer and Douglas N. C. Lin}
\affil{UCO/Lick Observatory, University of California, Santa Cruz,
CA 95064, U.S.A.}
\centerline{\today}

\begin{abstract}
Several short-period Jupiter-mass planets have been discovered around
nearby solar-type stars.  During the circularization of their orbits,
the dissipation of tidal disturbance by their host stars heats the
interior and inflates the sizes of these planets. Based on a series
of internal structure calculations for giant planets, we examine the
physical processes which determine their luminosity-radius
relation. In the gaseous envelope of these planets, efficient
convection enforces a nearly adiabatic stratification.
During their gravitational contraction, the planets' radii are
determined, through the condition of a quasi-hydrostatic equilibrium,
by their central pressure.  In interiors of mature, compact, distant
planets, such as Jupiter, degeneracy pressure and  the non-ideal equation
of state determine their structure. But, in order for young or
intensely heated gas giant planets to attain quasi-hydrostatic
equilibria, with sizes comparable to or larger than two Jupiter radii,
their interiors must have sufficiently high temperature and low density
such that degeneracy effects are relatively weak.
Consequently, the effective polytropic 
index monotonically increases whereas the central temperature
increases and then decreases with the planets' size.  These effects,
along with a temperature-sensitive opacity for the radiative surface
layers of giant planets, cause the power index of the luminosity's dependence
on radius  to decrease with increasing radius.  For
planets larger than twice  Jupiter's radius, this index is
sufficiently small that they become unstable to tidal inflation.  We
make comparisons between cases of uniform heating and cases in which
the heating is concentrated in various locations within the giant
planet.  Based on these results we suggest that accurate measurement
of the sizes of close-in young Jupiters can be used to probe their
internal structure under the influence of tidal heating.
\end{abstract}

\section{Introduction}
One of the surprising findings in the search for planetary systems
around other stars is the discovery of extrasolar planets with
periods down to 3 days (\cite{mq95}).  Nearly all planets with period
less than 7 days have nearly circular orbits.  In contrast, known
extrasolar planets with periods longer than 2-3 weeks, have nearly a
uniform eccentricity distribution.   The
shortest-period planets and their host stars induce tidal perturbations  on
each other.  When these disturbances are dissipated, angular
momentum is exchanged between the planets and their host stars, leading toward
both a spin synchronization and orbital circularization (\cite{ra96}).

\cite{BLM} (hereafter Paper I), considered the effect of tidal
dissipation of energy during the synchronization of these planets'
spin and the circularization of their orbits.  In that analysis, they
compute a series of numerical models for the interior structures of
weakly eccentric Jovian planets at constant orbital distances under
the influence of interior tidal heating and stellar irradiation. In
these previous calculations, the interior heating rate per unit mass
was imposed to be constant in time and uniformly distributed within
the planet.  Under these assumptions, they showed that Jovian planets
can be inflated to equilibrium sizes considerably larger than those 
deduced for gravitationally contracting and externally heated
planets. For the transiting planet around HD 209458, they suggested
that provided its dimensionless dissipation $Q$-value is comparable to
that inferred for Jupiter (\cite{yp81}), a small eccentricity ($e
\simeq 0.03$) would provide adequate tidal heating to inflate it to
its observed size (\cite{brown01}).  Since the orbital circulation time
scale is expected to be shorter than the life span of the planet, they
  also suggested that this eccentricity may be excited by another
planet with a longer orbital period.  The prediction of a small
eccentricity and the existence of another planet are consistent with
existing data (\cite{bll03}).

There are at least two other scenarios for the unexpectedly large size
of HD 209458b.  Heating by stellar irradiation reduces the temperature
gradient and the radiative flux in the outer layers of short-period
planets.  This process could significantly slow down the
Kelvin-Helmholtz contraction of the planet and explain the large size
(\cite{bur00}). However, even though the stellar flux onto the
planet's surface is 5 orders of magnitude larger than that released by
the gravitational contraction and cooling of its envelope, this
heating effect alone increases the radius of the planet by about 10\%,
not by 40\% as observed (\cite{gs02}).

An alternative source is the kinetic heating induced by the
dissipation of the gas flow in the atmosphere which occurs because of
the pressure gradient between the day and night sides (\cite{gs02}).
In order to account for the observed size of the planet, conversion of
only 1\% of the incident radiative flux may be needed, provided that
the dissipation of induced kinetic energy into heat occurs at
sufficiently deep layers (tens to 100 bars).  \cite{sg02} suggest that
the Coriolis force associated with a synchronously spinning planet may
induce the circulation to penetrate that far into the planet's
interior, and that dissipation could occur through, for example,
Kelvin-Helmholtz instability.  A follow-up analysis suggests that this
effect may be limited (\cite{blb03}; \cite{jl03}).

In order to distinguish between these three scenarios, the effect of
tidal heating for planets with modest to large eccentricities was
considered.  In a follow-up paper (Gu {\it et al.} 2003, hereafter
Paper II), we showed that the size ($R_p$) of compact Jupiter-mass
planets slowly increases with the tidal dissipation rate.  For
computational simplicity, we adopted a conventional equilibrium tidal
model which describes the planets' continuous structural adjustment in
order for them to maintain a state of quasi-hydrostatic equilibrium in
the varying gravitational potential of their orbital companion.  In
this prescription, a phase lag into the response is introduced to
represent the lag being proportional to the tidal forcing frequency
and attributable to the viscosity of the body.  The phase lag gives
rise to a net tidal torque and dissipation of energy and the
efficiency of the tidal dissipation can be parametrized, whatever its
origin, by a specific dissipation function or $Q$-value (quality
factor) (\cite{gs66}).  External perturbation can also induce
dynamical tidal responses through the excitation of g modes (Ioannou
\& Lindzen 1993a,b) or inertial waves (\cite{ol04})
which can be damped by viscous dissipation in the interior (\cite{gn77})
or radiative and nonlinear dissipation in the
atmosphere of the planet (\cite{ltl97}).

In both the prescription for equilibrium and dynamical responses, the
tidal dissipation rate is a rapidly increasing function of the
planets' radius.  But, their surface luminosity increases even faster
with $R_p$ such that, planets with relatively small eccentricities and
modest to long periods attain a state of thermal equilibrium in which
the radiative loss on their surface is balanced by the tidal
dissipation in their interior.  For planets with short periods and
modest to large eccentricities, the rate of interior heating is
sufficiently large that their $R_p$ may inflate to more than two
Jupiter radii.  In this limit, the surface luminosity of the planet
becomes a less sensitive function of its $R_p$ and the eccentricity
damping rate is smaller than the expansion rate of the planet so that
the increases in their surface cooling rate cannot keep pace with the
enhanced dissipation rate due to their inflated sizes.  These planets
are expected to undergo runaway inflation and mass loss.  We suggested
that the absence of ultra-short-period Jupiter-mass planets with $P<3$
days, which corresponds to an orbital semi-major axis $a$ of 0.04 AU,
may be due to mass loss through Roche-lobe overflow resulting from
such a tidal inflation instability. Other scenarios have been proposed
to explain the lack of Jupiter-mass planets with $P<3$ days:
truncation of inner part of a disk (\cite{lin96}; \cite{kl02}),
orbital migration due to the spin-orbit tidal interaction between the
close-in planet and the parent star (\cite{ra96}; \cite{ws02};
\cite{pr02}; \cite{jiang03}), and Roche-overflowing planets with the
help of disk-planet interaction but without tidal inflation
(\cite{tri98}).

In this contribution, we continue our investigation on the internal
structure of tidally heated short-period planets.  The main issues to
be examined here are: 1) how does tidal energy dissipation actually
lead to the expansion of the envelope? 2) how does the internal
structure of the planet depend on the distribution of their internal
tidal dissipation rate? and 3) what are the important physical effects
which determine the tidal inflation stability of the planets?  These
issues are important in determining the mass-radius relation of
short-period planets which is directly observable.

Structural adjustments may also modify the efficiency of dissipation
and the planets' Q-value for both equilibrium and dynamical tides.  In
the extended convective envelopes of gaseous giant planets and
low-mass stars, turbulence can lead to dissipation of the motion that
results from the continual adjustment of the equilibrium
tide. However, the turbulent viscosity estimated from the
mixing-length theory ought to be reduced by a frequency-dependent
factor owing to the fact that the convective turnover time scale is
usually much longer than the period of the tidal forcing. Based on the
present-day structure of Jupiter, Goldreich \& Nicholson (1977)
estimated $Q\approx5\times10^{13}$. However, within the intensely
heated (by tidal dissipation) short-period extra solar planets,
convection is expected to be more rigorous with higher frequencies
whereas their tidal forcing frequencies are smaller than that of
Jupiter.  The Q-value for the equilibrium tide within extra solar
planets is likely to much smaller than that within Jupiter.

For the dynamical response of short-period extra solar planets planet,
the forcing frequencies are typically comparable to their spin
frequencies but are small compared to their dynamical frequencies.
Convective regions of the planets support inertial waves, which
possess a dense or continuous frequency spectrum in the absence of
viscosity, while any radiative regions support generalized Hough waves
(Ioannou \& Lindzen 1993a, b).  Inertial waves provide a natural
avenue for efficient tidal dissipation in most cases of interest.  The
resulting value of $Q$ depends, in principle, in a highly erratic way
on the forcing frequency. Since the planets' spin frequency adjusts
with their $R_p$, which is a time-delayed function of the tidal
dissipation rate within them, the efficiency of tidal dissipation may
fluctuate while the overall evolution is determined by a frequency-average 
Q-value (\cite{ol04}).

In \S2, we briefly recapitulate the basic equations which determine
the quasi-static evolution of the planets' structure.  We show the
simulation results for inflated giant planets in the case of constant
internal heating per unit mass in \S3 and analyze the results, in the
Appendix, in terms of polytropic models which allow us to conclude that the
onset of the tidal runaway inflation instability is regulated by a
transition in the equation of state for the interior gas from a
partially degenerate/non-ideal state toward a more ideal-gas state. We
examine the dependence of planetary adjustment on different locations
of the energy dissipation in \S4.  Finally, we summarize the results
and discuss their observational implications in \S5.

\section{The Planetary Structure Equations and Numerical Methods}
The internal structure of an inflated Jupiter is constructed in this
paper with the same numerical scheme as that used in Paper I. The code
employs the tabulated equation of state and adiabatic gradient
$\nabla_{\rm ad}$ described by Saumon, Chabrier, \& Van Horn (1995)
and the values of opacities are derived from those provided by
Alexander \& Ferguson (1994). With this scheme, we compute
distributions of thermodynamical parameters as a function of radius
for a spherically symmetric planet (i.e. 1-D calculation). The surface
temperature of the planet is assumed to be maintained by the
irradiation from a solar-type star located 0.04 AU away. The structure
of the planet is assumed to be hydrostatic and not to be affected by
rotation since the nearly synchronized short-period planets spin at
least several times slower than Jupiter and their rotational energy is
quite small compared to their gravitational energy (paper
I)

The models include a tidal heating function, whose
physical basis is not well understood. Thus in
\S3 we consider simulations in which the planet
has a constant internal heating which is uniform
in mass. In \S4 we consider a set of models in
which the heating is non-uniform in mass. As
discussed by Ogilvie \& Lin (2004), different
mechanisms may operate in radiative or convective
layers. One should, in principle allow for three
avenues of tidal dissipation: viscous dissipation
of the equilibrium tide, viscous dissipation of
inertial waves, and emission of Hough waves in the
radiative zone.  Rotational effects can also affect
the behavior of tidal dissipation. In view of the uncertainties, we
do not model specific mechanisms but simply parameterize the heating rate.

In our calculations, we assume that convection
is so efficient that the temperature gradient behaves adiabatically
within a convective zone (Guillot  et al. 2003). In other words,
we solve the following equations for the radius $r$, the density
$\rho$, the pressure $P$, the temperature $T$, and the intrinsic
luminosity $\cal L$ (e.g. see \cite{kw})
\begin{eqnarray}
{\partial r \over \partial m}&=&{1\over 4\pi r^2 \rho},\label{eq:r} \\
{\partial P \over \partial m}&=&-{Gm \over 4\pi r^4}, \label{eq:P}\\
{\partial T \over \partial m}&=&-{GmT \over 4\pi r^4 P}\nabla, 
\label{eq:T}\\
\nabla&=&\nabla_{rad}={3\over 16\pi ac G}{\kappa {\cal L} P
\over mT^4}
\ \ {\rm if\ radiative}; \qquad
\nabla=\nabla_{\rm ad} \ \ {\rm if\ convective},\label{eq:nabla} \\
{\partial {\cal L} \over \partial m}&=&\epsilon - c_P
{\partial T \over \partial t} +{\delta \over \rho}
{\partial P \over \partial t},\label{eq:energy}
\end{eqnarray}
where $m$ is the Lagrangian mass coordinate, $\delta$ ($\equiv
-(\partial \ln \rho /\partial \ln T)_P$) is the coefficient of thermal
expansion at constant pressure, $c_P$ is the specific heat at constant
pressure, $\epsilon$ is the heating rate per unit mass, and $\nabla$
is defined as $d\ln T/d\ln P$.

The above structure equations are solved simultaneously with an 1-D
implicit Lagrangian scheme which uses the mass coordinate $m$ as an
independent variable.  In the case where the heating rate is
uniformly distributed with mass (i.e. $\epsilon=$constant), the
results of our numerical calculations usually show that the planet
interior is largely convective. Consequently, without the aid of the
energy equation (\ref{eq:energy}), equations (\ref{eq:r})
--(\ref{eq:nabla}) imply that the adiabatic assumption gives rise to a
unique radial stratification for a given planet's size and mass,
regardless of how strong the internal heating rate is or whether the
planet is in thermal equilibrium.  Equation (\ref{eq:energy}) then
indicates how fast the planet expands or contracts due to thermal
imbalance.  These adjustments proceed through a series of 
quasi-hydrostatic and quasi-thermal equilibria.

The initial condition is obtained from calculations
for the formation phases of planets of the appropriate
masses, as described in \cite{bhl00}.
Just after accretion ends, those calculations show that
the radius is about $2R_J$, which is the initial condition
used here. This value is somewhat uncertain, and it
changes quite rapidly during the earliest part of the
planet's cooling phase, but the exact value makes little
difference here, because most of the models reach a
thermal equilibrium which is practically independent of
the initial state.

For the inner boundary, we consider models with or without cores.
For the models with cores, we assume that they have a constant density
$\rho_{\rm core} = 5$ g cm$^{-3}$.  Their temperature is also assumed
to be the same as that of the envelope immediately outside the core.
The luminosity generated in the cores is assumed to be negligible.  In
most of our models, the central temperature is above $\sim 3\times
10^4$ K,  and the heavy elements in the cores are likely to be soluble
(Paper I).  Since we have already shown that core-less models lead to
more inflated planetary structure, the planetary radii determined for
the core models represent a lower limit.  There are some uncertainties in
the equation of state at the temperatures and pressures of the central
regions, where the hydrogen-helium gas is partially degenerate and
non-ideal.  We show below  that the degree of degeneracy and the non-ideal
effects are important in determining the stability against the tidal
runaway instability.

Near the planet's surface, the gas becomes radiative when $\nabla_r <
\nabla_{ad}$,  where $\nabla_r$ is computed using equation (\ref{eq:nabla})
and $\nabla_{ad}$ is obtained from a tabulated equation of state.  The
depth of the radiative zone is sensitive to the opacity.  The
temperature of the  surface layer is sufficiently low for grains
to condense.  We adopt the standard opacity table (\cite{af94}) which
is computed for the interstellar gas.  However, in Jupiter's
atmosphere  the  sedimentation of grains  leads to a much reduced opacity which
modifies the structure of the radiative zone.  For short-period
planets, however, the side facing the host star is heated to $\sim
1,500 $ K which is near or above the sublimation temperature of most
grain species.  In addition, a large scale circulation flow may also
modify the composition and the heat transport process near the
surface.  It remains to be determined whether grain sedimentation
occurs (Burkert  et al. 2003).  In order to take into account 
this possibility, we consider some models under the assumption of
extreme grain depletion, {\it i.e.} without any grain opacity,  so that
the effective opacity $\kappa_d$ is reduced from the normal opacity
$\kappa_0$ (see Table \ref{tab:model}).

At the surface of the planet, we include the irradiation from the star
in the boundary condition for the temperature $T_{surf}$.  We neglect
the difference between the day and night sides of the planet. For most
of the models, we set the irradiation temperature $T_0 = 1500$ K, which
corresponds to a planet with a 3-day period around a solar-type star.
For two models, we also considered somewhat lower temperatures.  At
the photosphere, the normal boundary conditions for pressure $P = 2 G
M_p / 3 \kappa_d R_p^2$ and for the intrinsic luminosity ${\cal L}=4
\pi \sigma R_p^2 (T_{surf}^4 - T_0^4)$ are adopted.

\section{Constant internal heating per unit mass}
In this section, we shall show the numerical results for an internal
heating rate that is constant in time and is uniform in mass; that is,
$\epsilon$ in equation (\ref{eq:energy}) takes   a constant value.  The
calculations are carried out for three different  masses:
$0.63M_J$, $1M_J$, and $3M_J$.
The time span for each calculation covers a few $10^9$ years. At the
end of each computation, the planets have attained  a thermal
equilibrium.
In this state, the integrated energy term $\epsilon$ in equation
(\ref{eq:energy})
balances the intrinsic radiated luminosity. Normally when a
planet contracts and cools, the time-dependent terms in
equation (\ref{eq:energy})
dominate, and a thermal balance is never reached.
The same is true even if the planet is strongly irradiated by
the star. However the present  results show that an internal
tidal energy dissipation rate on the order of $10^{27}$ to 
$10^{30}$
erg per second, depending on the mass, is sufficient to stop
the contraction and cooling at some radius in the range of a
few $R_J$.

In Figure \ref{fig:luminosity}, we show that there
exists a radius-luminosity relation for a Jovian planet with a given
mass $M_p$.  The planet's intrinsic luminosity (i.e.  excluding the
luminosity due to stellar irradiation) emitted from the photosphere
is denoted by $\cal L$. Discrete points shown in the Figure are our
computational results, which are fitted by three quadratic curves
associated with three different masses.  For a $0.63M_J$ core-less
planet, the solid curve in the left panel can be approximated by, 
\begin{equation}
\log ({\cal L}/10^{27}{\rm
erg/s})=-11.7039(\log (R_p/R_J))^2 +12.8994\log (R_p/R_J) -2.12471.
\end{equation}
For the same mass planet with a core (solid curve in the right panel),
we can approximate
\begin{equation}
\log ({\cal L}/10^{27}{\rm erg/s})=-8.78501(\log (R_p/R_J))^2
+10.004\log (R_p/R_J) -1.05358.
\end{equation}  
Similarly, the model without  a core (dashed curve in the left panel) and with
a core 
(dashed curve in the right panel) for a $1M_J$ planet can be approximated
by
\begin{equation}
\log ({\cal L}/10^{27}{\rm erg/s})=-7.7909(\log
(R_p/R_J))^2 +10.3338\log (R_p/R_J) -1.20809, 
\end{equation}
\begin{equation}
\log ({\cal L}/10^{27}{\rm
erg/s})=-6.74289(\log (R_p/R_J))^2 +8.92021\log (R_p/R_J) -0.573055
\end{equation}
respectively.  Finally, $3M_J$ models without a core  (dotted curve in the
left panel) and with a core  (dotted curve in the right panel) can be
approximated by
\begin{equation}
\log ({\cal L}/10^{27}{\rm erg/s})=-8.10772(\log (R_p/R_J))^2
+10.6893\log (R_p/R_J) -0.29812,
\end{equation}
\begin{equation}
\log ({\cal L}/10^{27}{\rm erg/s})=-10.525(\log
(R_p/R_J))^2 +12.3851\log (R_p/R_J) -0.508571
\end{equation} respectively.  

For large values of $R_p$, these three curves all become flattened as
the planet is inflated; {\it i.e.} as the planet expands, the slope
($\gamma \equiv d\log {\cal L}/d\log R_p$) decreases. In the core-less
cases, $\gamma= 5$ at $R_p=2.175 R_J$ for $M_p=0.63M_J$,
$R_p=2.199R_J$ for $M_p=1M_J$, and $R_p=2.243R_J$ for $M_p=3M_J$.  In
the core cases, $\gamma= 5$ at $R_p=1.927 R_J$ for $M_p=0.63M_J$,
$R_p=1.953R_J$ for $M_p=1M_J$, and $R_p=2.243R_J$ for $M_p=3M_J$.
Also, 
the luminosity for a given $R_p$ is an increasing function of $M_p$.

The results in Figure \ref{fig:luminosity} do not correspond to the
exact thermal equilibrium solutions.  Although the planets quickly
establish a hydrostatic equilibrium, ${\cal L}$ does not necessarily
equal the internal heating rate.  In these calculations, ${\cal L}$
is the instantaneous intrinsic luminosity associated with the
corresponding $R_p$.  We use various initial conditions for $R_p$ and
$\epsilon$ for our computation, but the ${\cal L}$ vs $R_p$ curves
remain nearly identical to those shown in Figure~\ref{fig:luminosity}.
This invariance is due to the  adiabatic structure as we
have stated in the previous section. The number marked on each data
point is the degree of electron degeneracy at the center of the planet
in the core-less cases or at the surface of the core in the core cases.
The degree of electron degeneracy is expressed by the ratio of the
Fermi energy to the thermal energy 
\begin{equation}
D\equiv E_{Fermi}/kT \approx \rho/ (6\times 10^{-9} T^{3/2}). 
\label{eq:D}
\end{equation}
As shown in the figure, the degeneracy is lifted as the planet
expands.

In Figure \ref{fig:nabla}, we plot the temperature against the
pressure inside a planet with $M_p=1M_J$. Each curve corresponds to a
different value of $R_p$. While the upper-right parts of the curves,
where $T$ and $P$ are higher, represent the conditions for planet
interiors, the lower-left portions of the  curves show the $T$ and $P$
distribution near the planet's surface.  The thick line across the  upper
parts of the curves marks the ``Plasma Phase Transition (PPT)'' line
which segregates hydrogen molecules and metallic hydrogen atoms.  This
line is approximately drawn based upon an extrapolation of a hydrogen
phase diagram (\cite{SCV}).  As indicated by the slope of the curves,
$\nabla_{ad}$ has a value $\simeq 1/3$ in the interiors where
hydrogen atoms are in metallic form, and then drops down to around 0.2
as we follow the curves to the outer region of the planet where the
pressure decreases below 10 bars.  This transition occurs near the
base of the radiative envelope. The flattening of $\nabla$ in the
radiative envelope results from the external heating due to the
stellar irradiation. The overall behavior of the curves for core and
for core-less cases is very similar. In the core-less case, a decrease
in the central pressure $P_c$ with the planet's size can be fitted
with a following power law
\begin{equation}
P_c \propto R_p^{-2.24}.\label{eq:P_c}
\end{equation}
The above relation is less steep than what is expected from a purely
polytropic structure in which $P_c \propto R_p^{-4}$. In addition, the
central temperature $T_c$ increases with $R_p$ when $R_p \lesssim 2.3
R_J$, but decreases with $R_p$ when $R_p \gtrsim 2.3 R_J$. This
variation of temperature is nearly independent of the strength of
internal heating rates; that is, the variation of temperature profiles
is almost the same for a given planet's size no matter whether the
planet is expanding, contracting or even in a thermal equilibrium.
This tendency arises because, as stated in the previous section,
adiabatic profiles are imposed in the convection zone which amounts to
95\% of the total radius. The change of internal temperature in
this manner is not
driven by a thermal imbalance, but it is caused by an intrinsic change
in the physical properties of the fluid in the convection zone as
a result of the reduction of electron degeneracy (see Appendix A.2).

Figure \ref{fig:degeneracy} displays the mass density profiles as a
function of the temperature for various planet sizes. The
theoretical Plasma Phase Transition line is also marked in the
plot. Resembling the flattened $P-T$ curves in Figure \ref{fig:nabla},
the steepening of the $T-\rho$ curves for $\rho$ less than about
$10^{-4}$ g/cm$^3$, which roughly corresponds to the region of
radiative envelope, is caused by the stellar irradiation. We estimate the
degree of degeneracy at the center of the planet with no core, and at
the surface of the core for the planet with a core.  As the planet
with no core (with a core) expands from $1.777R_J$ to $2.826R_J$
($1.6R_J$ to $3.9R_J$), the value $D$ declines from 21 to 9 (from 21
to 5.3) due primarily to the central density decrease.  We fit the
relation between $\rho_c$ and $R_p$ for the planet with no core by
using the following power law
\begin{equation}
\rho_c \propto R_p^{-1.6}.\label{eq:rho_c}
\end{equation}
 As is the case for the $R_p-P_c$ relation, which is described by
equation (\ref{eq:P_c}), the power 1.6 in equation (\ref{eq:rho_c}) is
smaller than what is expected from a purely polytropic structure in
which the power is 3.

The interior model of Jupiter is usually approximated by an $n=1$
polytrope, which results from the combination of Coulomb effects and
electron degeneracy in the metallic hydrogen plasma (\cite{hubbard};
\cite{bhll}).  This simplification is confirmed by the simulation as
shown in Figure \ref{fig:polytrope} which plots $P/\rho^2$ (in c.g.s
units) as a function of the co-moving mass coordinate $m$ for various
planet sizes.  The hypothetical plasma phase transition line is
denoted by PPT and is marked by a thick line. The almost constant
value of $P/\rho^2$ throughout the region of metallic hydrogen
indicates a $n\approx 1$ polytropic structure. As we follow these
curves to the  outer regions of the planet where the molecular hydrogen
dominates, the slopes of these curves increase from zero, meaning that
the polytropic index is increased.  The fractional region of the $n=1$
polytropic structure shrinks as the planet's size becomes larger. This
tendency suggests that the interior structure at various evolutionary
stages of an inflated planet does not proceed in a self-similar
manner, but behaves differentially as a consequence of the evolution of the
equation of state. This motivates us to employ the simple
polytrope approach to sketch the physical properties of an inflated
planet associated with different values of the polytropic index $n$.
By fitting the curves in the region of convection zones in Figure
\ref{fig:polytrope} for the core-less cases with a polytropic index,
we show in Table \ref{tab:polytrope} that the polytropic index increases
with $R_p$. As demonstrated by the polytropic approach compared to
the simulations in Appendix, 
the increase in $n$, and
therefore the decrease in the adiabatic exponent
$\Gamma$ (see equation (\ref{eq:n_delta})
and the right panel in Figure \ref{fig:delta} in Appendix),
with $R_p$ for the interior
structure arises from a migration of an equation of state
from a more degenerate and non-ideal phase
to a state with less degenerate and more ideal
properties. Consequently, for a given planet mass,
the rate of $R_p$ change in response
to the change of total entropy becomes more sensitive as $n$
increases (see equation(\ref{eq:C_M}) for the planet compressibility
$C_M$ and the description in Appendix); 
namely, a larger planet is more elastic with
respect to the change of entropy and pressure. This
explains the simulation results that the increase in
the intrinsic luminosity $\cal L$ is reduced as the planet
expands as shown 
in Figure \ref{fig:luminosity} and equation (\ref{eq:gamma_Rp}),
making a
larger planet vulnerable to the tidal inflation instability
if the dimensionless parameter for the tidal dissipation $Q$ 
remains unchanged during the inflation.

Although the above analysis is based on the simulation for
uniform heating rate per unit mass, the concept of planet
elasticity depending on the planet's size can be
extended to the off-center heating cases.
This will be discussed in the next section.

\section{Dependence on the location of the tidal dissipation}

\subsection{Prescription for tidal dissipation rate}
In general, the distribution of tidal energy dissipation is a function
of the radius, {\it i.e.} $\epsilon = \epsilon (r)$.  As we indicated
above, the actual functional form $\epsilon (r)$ depends not only
the location of wave excitation and propagation, but also on the nature
of dissipation.  In stars and gaseous planets, the tidal perturbation
of a companion not only distort their equilibrium shape but also
excite gravity and inertial waves (Zahn 1977, 1989; Ioannou \& Lindzen
1993a,b, 1994, Ogilvie \& Lin 2004).  The equilibrium adjustment and
the dynamical waves which propagate within the planets are dissipated
either through convective turbulence in its interior or radiative
damping near the surface (Zahn 1977, 1989, Goldreich \& Nicholson
1977, Papaloizou \& Savonije 1985, Xiong {\it et al.}  1997, Lubow et
al. 1997; Terquem et al. 1998; Goodman \& Oh 1997). Tidal dissipation
of the equilibrium tide deep inside a convective giant planet might be
small because the convective flow is so adiabatic that the eddy
turnover time of convection is much larger than the tidal forcing
period which is about 3 days for the close-in planets we are
investigating, leading to the outcome that the dissipation is
restricted to an area far from the center of the planet
interior. However, the tidal dissipation might take place at greater
rates than the conventional calculations estimate through resonance
locking between the perturbing tidal potential and oscillation
(\cite{ws02}), which might happen to the tidal dissipation inside a
giant planet. Most importantly in the case of a convective hot
Jupiter, the dissipation rate associated with inertial waves in
resonance with the harmonics of tidal forcing might not be severely
reduced in the regime of low viscosity because the dissipation is
independent of the viscosity (\cite{ol04}).  The dissipation of the
waves, which occurs throughout the planets' convective envelope, also
leads to the deposition of angular momentum which in general leads to
differential rotation (Goldreich \& Nicholson 1977, 1989).  It is not
clear how the interior of the gaseous planets and stars may readjust
after the angular momentum deposition may have introduced a
differential rotation in their interior and how the energy stored in
the shear may be dissipated (\cite{kpb91}).

In this paper, we are focused on the response of the planets' envelope
to the release of energy associated with the tidal dissipation.  To
determine the actual dissipation distribution is beyond the scope of
the present work. In light of these uncertainties, we adopt three {\it
ad hoc} prescriptions for the positional dependence in $\epsilon (r)$
such that
\begin{equation}
\epsilon_1 (m) = \left( {(1 + \beta) \epsilon_0 \over M_p} \right)
\left( {m \over M_p} \right)^\beta
\end{equation}
\begin{equation}
\epsilon_2 (m) = \left( {(1 + \beta) \epsilon_0 \over M_p} \right)
\left( {M_p - m \over M_p} \right)^\beta
\end{equation}
\begin{equation}
\epsilon_3 (m) = \left( {2 \epsilon_0 \over {\sqrt \pi }M_p} \right)
{\rm exp} \left( - {(m - m_0)^2 \over \Delta m^2} \right)
\end{equation}
where $M_p$ is the planets' total mass, $\beta$ is a power index,
$m_0$ and $\Delta m$ determine the Gaussian profile.  For $\beta>0$,
$\epsilon_1$ corresponds to a concentrated dissipation near the
planet's surface such as that due to the atmospheric nonlinear
dissipation or radiative damping, $\epsilon_2$ corresponds to intense
dissipation within the planets' envelope such as that expected from
turbulent dissipation and nonlinear shock of resonant inertial waves,
whereas $\epsilon_3$ is designed to high light the dissipation near
some transition zone such as the convective radiative transition
front, or near the corotation radius.
In the notion of equilibrium tides with constant lag angle, the
total tidal energy dissipation rate within the planet in its rest
frame (\cite{egg98}, \cite{ml02}, papers I and II) is
\begin{equation}
\epsilon_0 = \dot E_{\rm tide} =- \left({9 \mu n a^2 \over 2 Q_p^\prime}
\right) \left({M_\ast \over M_p} \right) \left({R_p \over a} \right)^5
\left[ \Omega_p^2 h_3(e)- 2 n\Omega_p h_4(e) +
n^2 h_5(e) \right]
\label{eq:etide24},
\end{equation}
where $M_\ast$ and $M_p$ are the mass of the stars and their planets,
$\mu= M_\ast M_p/(M_\ast + M_p)$ is the reduced mass, $\Omega_p$ and
$R_p$ are the planets' spin frequency and size, $n$, $a$, $e$ are the
mean motion, semi major axis, and eccentricity of the planets' orbit,
$h_3(e)=(1+3e^2 + 3e^4/8) (1-e^2)^{-9/2}$, $h_4(e)=(1 + 15 e^2/2 +45
e^4/8 + 5 e^6/16) (1 - e^2)^{-6}$, and $h_5(e) =(1+31e^2/2+255e^4/8
+185e^6/16+25e^8/64) (1 - e^2)^{-15/2}$. In the limit of low
eccentricities, equation(\ref{eq:etide24}) can be approximated
as follows:
\begin{eqnarray}
\epsilon_0 &\approx &{e^2 GM_* M_p \over a \tau_d} \nonumber \\
&\approx& 6.1 \times 10^{29} e^2
\left( {M_* \over \msun} \right)
\left( {M_p \over M_J} \right)
\left( {10\rsun \over a} \right)
\left( {20 {\rm Myr} \over \tau_d} \right) \ {\rm ergs\,s^{-1}},
\label{eq:epsilon_0}
\end{eqnarray}
where
\begin{equation}
\tau_d \approx 20
\left( {Q_p \over 10^6} \right)
\left( {M_p \over M_J} \right)
\left( {\msun \over M_*} \right)^{3/2}
\left( {a \over 10\rsun} \right)^{13/2}
\left( {2R_J \over R_p} \right)^5 \ {\rm Myr},
\label{eq:tau_d}
\end{equation}
and $Q_p$ is the dimensionless parameter which
quantifies tidal deformation and dissipation of a planet (paper II).

An equilibrium tidal model with a constant lag angle for all
components of the tide may not necessarily be the most reliable model.
All the uncertainties associated with the physical processes are
contained in the $Q_p$ values and we shall parameterize our results in
terms of it.  For a fiducial value, we note that the $Q_p$-value
inferred for Jupiter from Io's orbital evolution is $5 \times 10^4 <
Q_p < 2 \times 10^6$ (\cite{yp81}).  With this $Q_p$-value, orbits of
planets with $M_p$ and $R_p$ comparable to those of Jupiter, and with
a period less than a week, are circularized within the main-sequence
life span of solar-type stars as observed.

For identical forcing to spin frequency ratios $f_\Omega$, the magnitude of
$\epsilon_0$ for dynamical tides has the same power law dependence on
$R_p$ as that for equilibrium tide in eq (\ref{eq:etide24}).  But,
the dissipation rate and the Q-value vary sensitively with $f_\Omega$
which is modulated by the changes in $R_p$ (Ogilvie \& Lin 2004).
Numerical calculation and analytic approximation show that the relevant
frequency-averaged Q-value is comparable to that inferred for Jupiter
and it may be asymptotically independent of the viscosity in the limit
of small viscosity or equivalently Ekman number.

In the present context, we assume $\epsilon_0$ is constant in time in
most of our models. But, for
models 18--20, we consider the possibility that the damping time scale
for the eccentricity
($\tau_d$) is longer than the thermal expansion time scale ($\tau_R$)
for the planets to inflate. 
The expression for $\tau_R$ is given by
the equation (paper II)
\begin{equation}
\tau_R ={e_R^2 \over e^2 }\tau_d,\ {\rm where}\ 
e_R=\left( {q_p M_p a \over \beta M_* R_P} \right)^{1/2}
\approx 0.18
\left( {q_p\over 0.75\beta}{M_p \over M_J}{\msun \over M_*}
{2R_J \over R_p}{a\over 0.04\,{\rm AU}} \right)^{1/2}.
\label{eq:e_R}
\end{equation}
Hence in these models the heating rate  
$\epsilon_0$ increases as $R_p^5$ according to equation (\ref{eq:etide24})
with a constant eccentricity. 

\subsection{Numerical models}
In order to explore the dominant effect of tidal inflation, we
consider several models (see Table \ref{tab:model}) for an 1$M_J$
Jupiter being inflated from the same initial  size
of  1.9 $R_J$. The parameters for these
models are chosen to represent a wide range of possibilities.

In the first and second series of models, mild (strong) heating is
deposited in different locations in models 1--8 (9--12).  With models
13-15 in the third series, we consider the possibility of smaller
opacities, due to grains' sedimentation, in the radiative envelope.
With a lower $T_e$, we consider the possibility of less intense
stellar irradiation in models 16  and  17 in the fourth series.  Simple
forms of time-varying tidal dissipation are also considered in the
fifth series of models 18--20. We consider
the damping timescale of eccentricity $\tau_d$ is longer than the
expansion time scale, so the heating rate in models 18--20
takes the tidal form $\epsilon \propto R_p^5$.
We only deal with the core-less cases
for all of these models shown in Table \ref{tab:model}
except for model 19 in which the simulations for
both core and core-less models are carried out.

\subsubsection{Dependences on the heating location}
In this series of models, we consider the effect of non-uniform
heating on the planets' radius-luminosity relation. The relation which
stays approximately invariant for different locations of heat deposit
is the radius-adiabat relation (\cite{stahler}). In Figure
\ref{fig:rhoModels}, we plot the density profiles of an $ 1 M_J$ inflated giant
planet in a thermal equilibrium.  Model 1 \& 6 are
represented by the dotted line,  because their interior structure is
almost the same for these two models as the heating is concentrated in
the inner regions of the planet.  Models 2 and 5 are represented by
solid and dashed lines respectively.  

These models indicate that as the location of maximal heating moves
from the planet's center toward the photosphere ($m_0/M_J=0.05
\rightarrow 0.7 \rightarrow 0.9999$), the final equilibrium size of
the planet decreases.  This tendency indicates that less $PdV$ work on
expansion is achieved when the heating shell is closer to the
photosphere because the heating is also more efficiently lost near the
planet's surface. Equivalently, less specific entropy content is
retained by the convective region of the planet. Given a mass, a size, 
and the same  heating rate, a planet with the heating
shell closer to its photosphere has larger intrinsic luminosity $\cal
L$ as a result of the difficulty in transporting entropy into the
planet's interior via both radiative and convective transport. Our
numerical results indicate that the region beneath the main heating
zone continues to adjust slowly without gaining any entropy. Even
without a substantial increase in entropy, we show in the next
paragraph for models 11 and 12, that the inner convective region can
swell in order to adjust to a new hydrostatic equilibrium.  This
adjustment is due to a drop in the boundary pressure in the limit
that the spacial extent of the heating shell spreads significantly.
Although this result seems to be in qualitative agreement with the
radius-adiabat relation, it is still unclear that the relation should
precisely hold for a giant planet heated in different locations
because different equations of state and therefore different $PdV$
work are involved for a given entropy input to different locations.

Figure \ref{fig:case11+12} displays the evolution of expansion for
model 11 (upper panels) and model 12 (low panels). The three solid
curves from right to left in each of the temperature and density
diagrams are the profiles corresponding to the phases marked by the
data points 1, 6, 11 in the $R_p$ vs $t$ diagram for model 11, and the
phases labeled by the data points 1, 4, 10 in the $R_p$ vs $t$ diagram
for model 12.  The vertical dashed lines in the temperature and
density diagrams denote the location of the center of the Gaussian
heating zones: $m/M_p=0.9$ for model 11, and $m/M_p=0.7$ for model 12. In
model 11, a heating front is generated at $m/M_p=0.9$ as is illustrated
in the temperature diagram (i.e. the upper-middle panel) and then
propagates inward. The front corresponds to the region with the
temperature inversion. It also acts as a rarefaction structure as is
shown in the density diagram. For comparison, we also draw the
temperature and density profiles for model 9 (dot-dashed curves) and
for model 10 (dotted curves). In model 12, a heating front is
generated at $m/M_p=0.7$. As it travels inward (see the lower-middle
panel), the front decreases the density analogous to a rarefaction
wave (see the lower-right panel).  

It is interesting to point out that except for the data points from 1
to 4 in model 12, all of the other data points in the $R_p$ versus $t$
diagrams correspond to a state of quasi-thermal equilibrium; i.e. the
intrinsic luminosity $\cal L$ is equal to the dissipation rate. 
However the planet is not in a global thermal
equilibrium because of the presence of a temperature inversion
in the region just below the shell. This heat front very
gradually propagates inward, and is associated with a gradual
overall expansion. After quasi-thermal equilibrium is reached,
the rate of expansion is very slow compared to its rate  before
this time (compare the data before and after point 4 in the lower
left panel of Figure \ref{fig:case11+12}). Clearly, an analogous expansion 
phase does not
occur in the uniformly heated case (model 9).

The underlying physical cause for expansion of $R_p$ without any gain
of internal entropy (or very little gain of entropy as a result of the
tail of a Gaussian profile) can be described as follows. Prior to
reaching a quasi-thermal equilibrium, a giant planet experiences a
fast inflation due to a high dissipation rate focused in a spherical
envelope (such as the stage before the data point 4 for model 12 in
Figure \ref{fig:case11+12}). In models 11 and 12, about 10\% to 30\%
of planet gas by mass lies above the heating zone.  This gas is
inflated significantly due to a strong underlying shell-heating source
as prescribed by the model. The huge inflation on the top of the
heating shell reduces the pressure and therefore causes the whole
region beneath the heating zone to expand and to adjust to a new
hydrostatic equilibrium. 
The expansion induced by a shell-heating source is less efficient than
that induced by a uniform heating throughout the envelope since more
energy is radiated away for a given $R_p$, as we have already stated
in the previous paragraph. The surface radiation becomes more
intensified as the heating zone approaches to the photosphere.  

After
the entire shell-heated planet has reached its quasi-thermal
equilibrium, the region around the heating front is not yet in the
thermal equilibrium locally. This arises because the region
just below the heating front (i.e. at the bottom of the temperature
inversion region) gains entropy via convection from the
bottom and through radiation from its top.  But the region just above
the heating front (i.e. at the top of the temperature inversion region)
loses entropy due to radiation from its bottom and
convection from its top.  On the average, the net increase in entropy
vanishes.  Thus, the rarefaction heating wave is driven by a local
entropy transport which averages to a zero net flux. Since the heating
front is a result of the temperature inversion which transports
entropy inward by radiative diffusion and causes the entropy
disturbance, the time scale for the heating front to cross the
planet's interior is roughly equal to
\begin{eqnarray}
t_{rarefaction} &\sim& {R_p \over \nu_{rad} / \Delta r} \nonumber \\
&\approx& 40
\left( {R_p \over 3 R_J} \right)
\left( {\Delta r \over 10^8\,{\rm cm}} \right)
\left( {3.5\times 10^4\,{\rm K} \over T} \right)^3 
\left( {\kappa \over 10^3\,{\rm cm^2/g} } \right) 
\left( { \rho \over 0.3\,{\rm g/cm^3}} \right)^2
\left( { c_p \over {\cal R}} \right) \ {\rm Gys},
\end{eqnarray}
where the radiative diffusivity $\nu_{rad}=4ac T^3/3\kappa \rho^2
c_p$, and $\Delta r$ is the width of the heating front.  The reference
values for $\Delta r$, $T$, $\kappa$, and $\rho$ shown in the above
estimate are taken from the simulation for model 12.  The large value
of $t_{rarefaction}$ is consistent with our numerical results that the
shell-heating planet expands very slowly after it has reached the
quasi-thermal equilibrium (see the upper-left panel and the stage
after the data point 4 in the lower-left panel in Figure
\ref{fig:case11+12}). In reality the slow expansion of the planet
due to the inward propagation of the rarefaction wave is
unimportant, because the damping time for
tidal dissipation is much short than $t_{rarefaction}$, and
also $t_{rarefaction}$ is much longer than the lifetime of the star.

The ${\cal L}-R_p$ relation for model 12 before the planet reaches a
quasi-thermal equilibrium can be fitted with a quadratic
approximation:
\begin{equation}
\log ({\cal L}/10^{27}\,{\rm erg/s}) =-9.36888(\log
R_p/R_J)^2+11.9403 (\log R_p/R_J) -1.33544, 
\end{equation}
which indicates that $\gamma$ decreases with $R_p$ and $\gamma= 5$
when $R_p=2.346R_J$. In contrast to the uniform heating cases, in
which the planet's interior gains entropy and expands, the degree of
degeneracy at the planet's center, in model 12, remains essentially
constant ($D\approx 17$) during the evolution (see Figure
\ref{fig:case11+12}). This tendency arises largely because the planet
interior expands without gaining much entropy. This result implies
that the flattening of $\cal L$ with $R_p$ in the shell-heating cases
is not due to the reduction of electron degeneracy of the metallic
interior, but is primarily caused by the gradual phase transformation
from non-ideal properties to the ideal state within and above the
shell-heating zone. This interpretation is well illustrated in
Figure~\ref{fig:Jdelta_12} in terms of $J\delta$ 
($=1/\Gamma$, see equation (\ref{eq:n_delta}) and the description in Appendix)
for the data point 1 (solid curve) and the data point 4 (dotted
curve): while in the interior $J\delta$ hardly changes    with
$R_p$, $J\delta$ indeed increases with $R_p$ around and above the
shell-heating zone.

\subsubsection{Variations in other model parameters}
In addition to the cases in which most of the dissipation is
concentrated in a narrow spherical shell, we also considered the cases
with relatively flat heating profiles for $\epsilon_0$.  In Model 7
and 8, we choose the power index to be $\beta=2$. We found that the
interior structure is almost the same for Models 7, 8, and 1. This
similarity arises because the heating per unit volume $\rho
\epsilon_0$ concentrated in the high-density inner region of the
planet in all these models, including Model 7 with $\beta=2$.  For
much larger values of $\beta$, the internal structure of Model 7 and 8
should resemble that of Model 2 and 6 respectively.

Figure~\ref{fig:kappa} depicts the density profiles of the planet in a
thermal equilibrium with the regular and the reduced opacities for
grains. The solid and dashed curves represent Models 1 and 4
respectively.  In both models, $\kappa_d=\kappa_0$.  The corresponding
Models 13 and 14 for $\kappa_d=10^{-3} \kappa_0$ are marked with solid
and dashed curves respectively.  This reduction of opacity by a factor
of $10^3$ is implemented in the grain region where $T<2100$ K. This
region roughly corresponds to the radiative envelope of the
planet. Since the radiative envelope is a bottleneck for the outward
energy transport, the reduced opacity in the radiative envelope allows
photons to escape from the planet much more  easily.  Consequently, the
planet attains a much smaller size when a thermal equilibrium is
established. The dashed curves have slightly smaller $R_p$ than the
solid curves as a result of more radiative loss for the
shell-heating cases as reasoned in the previous paragraph. Since
the location of the heating shell is close to the planet's center, the
simulation for Model 15 is almost the same as that for Model 13.

We also computed the thermal equilibrium solution for models with
different irradiation temperatures $T_e$.  In Figure~\ref{fig:Te}, we plot
the temperature and opacity distributions for Models 1, 16, and 17
with solid, dashed, and dotted curves respectively.  Since the
interior structures for these three cases are almost the same, we only
show the temperature and opacity profiles in the radiative envelopes,
which are  affected by the stellar irradiation.  The planets with less
irradiation (Models 16 \& 17) have slightly larger equilibrium sizes
than the one with more irradiation (Model 1).  A direct comparison
between Models 16 and 17 indicates that higher surface temperatures
result not only in a slight increase in the density scale height of
the planet's atmosphere, but also an enhanced opacity.  Both effects
cause Model 16 to attain a larger $R_p$ than Model 17.

But $R_p$ for Model 1 is smaller than that for Model 16 even though
the photospheric temperature for the former is higher than that of the
latter.  This difference arises  because the  temperature of the envelope in Model 1
is sufficiently high for most of the silicate grains to sublimate.
Consequently, opacity  in the radiative envelope in Model 1 is well
below that in Model 16.  The correlation between $R_p$ and $\kappa$ is
already established in the previous discussion on Models  13--15.
However, the difference in the size between Model 1 and Model 16 is
only about 1\%, in accord with the usual notion that the size of an
optically-thick object should not be strongly altered by external
irradiation.

\subsubsection{Self consistent tidal heating models and inflation 
instability}
We also consider, in Model 18, a self consistent calculation in which
the energy dissipation is proportional to $R_p^5$ in accordance with
equation (\ref{eq:etide24}).  In this model, we assume that the energy
dissipation is uniformly distributed in mass. The dissipation is
normalized to $\epsilon_0=5\times 10^{-6}\Lsun$ at 2 Jupiter radii
(this rate corresponds to $a=10\rsun$ and $e\approx 0.18$ in
equation (\ref{eq:epsilon_0}). Note that $\tau_R \sim \tau_d$ in this
case according to equation(\ref{eq:e_R})). The final equilibrium size
$R_p=1.857~R_J$ is slightly smaller than that ($R_p=1.977~R_J$) for
Model 1.  The planet would be slightly bigger ($R_p=1.983~R_J$) than
that for Model 1 if the normalization constant were increased by 5\%.
Since $\gamma<5$ as $R_p \gtrsim 2R_J$, the planet is expected to be
vulnerable to the tidal inflation instability once its size expands
beyond $2R_J$. Our numerical results show that a further increase in
the normalization constant by another 5\% (therefore $e\sim 0.2$ and
$\tau_R \sim 0.9 \tau_d$ when $R_p=2R_J$ and $a=10\rsun$ by
equation(\ref{eq:e_R})) causes $R_p$ to increase to at least 4 $R_J$.

In comparison to the uniform heating in Model 18, we consider in
Models 19 and 20 the possibility of non-uniform heating with $m=0.9$
and $m=0.9999$, respectively.  Similar to Model 18, the total heating
rate is proportional to $R_p^5$.  In Model 19 with the heating rate
normalized to 2.1 times larger than $5\times 10^{-6} \Lsun$ at
$R_p=2R_J$ (this corresponds to $e\approx 0.26$ at $a=10\rsun$
according to equation (\ref{eq:epsilon_0})), the planet is thermally
unstable and swells from 2$R_J$ to 3$R_J$ in about 2 Myrs (shorter
than the eccentricity damping time scale $\tau_d \approx 2.7$ Myrs
according to equation (\ref{eq:tau_d}) evaluated at $R_p=3R_J$).  When
the heating rate is normalized to 4 times larger than $5\times 10^{-6}
\Lsun$ at $R_p=2R_J$ (this corresponds to $e\approx 0.07$ at $a=0.03$
AU according to equation(\ref{eq:epsilon_0})) in Model 19, the planet
is thermally unstable and expands from 2$R_J$ to 3$R_J$ in about 0.27
Myrs (the eccentricity damping time scale $\tau_d \approx 0.15$ Myrs
and 1.17 Myrs evaluated at $R_p=3R_J$ and $R_p=2R_J$, respectively).

When the normalized heating is set to be $10^{-3}\Lsun$ for $a=0.03$
AU and $R_p=2R_J$, equation (\ref{eq:epsilon_0}) is no longer a fair
approximation and equation (27) in paper II gives $e\approx 0.294$ for
this heating rate.  In this case the planet with or without a core
expands from $2~R_J$ to $3.6~R_J$ in less than 40000 years which is
shorter than the eccentricity damping time $\approx 47800$ years for
$R_p=3.6~R_J$ according to equation (12) in paper II. Since the planet's
size $3.6~R_J$ is actually beyond the Roche radius $R_L\approx 3~R_J$
for $a=0.03$ AU and $e=0.294$, the inflated planet would overflow
the inner Lagrangian point in this case.  

The expansion rate is drastically reduced for Model 20 in which the
dissipation is largely deposited at $m=0.9999$, roughly the location
of the radiation-convection boundary.  A one-Jupiter mass planet with
the normalized heating $5\times 10^{-6} \Lsun$ at $R_p=2R_J$ can only
reach the final size of $\approx 1.9~R_J$. A normalized heating
rate which is twice as big as $5\times 10^{-6} \Lsun$ results in a
final size $\approx 1.95~R_J$. The planet with the normalized heating
four times as large as $5\times 10^{-6} \Lsun$ in Model 20 expands
very slowly to $2~R_J$ from $1.9~R_J$ over several tens of million
years, which is comparable with the eccentricity damping time scale
20 Myrs at $a=10\rsun$ and is much longer than the eccentricity damping
time scale ($\simeq$ 1.17 Myrs) at $a=0.03$ AU.  The planet in Model
20 obviously requires a larger normalized heating rate than Model 19
to reach the critical size beyond which the planet is thermally
unstable in response to the $R_p^5$ heating rate.

\section{Summary and discussion}
In this paper, we continue our investigation on the adjustment of
a planetary interior as a consequence of intense tidal heating.  As a giant planet's
interior is heated and inflated, we showed in Paper II that its
interior remains mostly convective.  Efficient energy transport leads
to an adiabatic stratification.  With a constant heating rate per unit
mass, we deduced an unique luminosity-radius relation regardless of
how intense the heating rate is.  The planet's luminosity increases
with its radius.  But the growth rate of $\cal L$ is a decreasing
function of $R_p$.  At the same time, the tidal dissipation 
heats the interior of the planet at a rate which  increases rapidly with $R_p$.  
At around $2~R_J$, $\cal L$ can no longer sustain sufficient growth 
to maintain a thermal equilibrium with the tidal dissipation rate.  
Thereafter, the planet's inflation become unstable and it overflows 
its Roche radius and become tidally disrupted.  

Here we show that the change of luminosity during the planet's
expansion is directly linked to the evolution of its interior,
in particular, the equation of state.  We employ the polytrope approach to
investigate the interior structure of an inflated giant planet.
According to the simulation, interior profiles deviate away from
$P\propto \rho^2$ as the planet expands. The central temperature
$T_c$ increases and then decreases with the size of the planet. Also
$P_c$ and $\rho_c$ are less steep functions of $R_p$ than the
polytrope theory with a constant polytropic  index $n$  indicates. All
of these effects 
suggest that the planet interior does not evolve in a self-similar
manner,  but $n$ gets larger  as $R_p$ increases.  In conjunction with the
numerical results that the degeneracy $D$ decreases, and that $T_c$
rises and then drops during the course of inflation, the result of a
positive value of $dn/dR_p$ can be interpreted as a manifestation of a
reduction in degeneracy during the expansion.  

We reason that the coefficient of thermal expansion $\delta$ increases
in response to a decrease in degeneracy and non-ideal effects, leading
to an increase in $n$ through equation (\ref{eq:n_delta}). Consequently
the planet compressibility at constant mass $C_M$ increases with $R_p$
(see equation \ref{eq:C_M}).  
This pattern can be translated into the
phenomenon of a decrease in luminosity growth during the inflation, as
a consequence of an one-to-one relation between $K$ and $\cal L$ in
the case of uniform heating in mass under the conditions of the
polytropic interior and hydrostatic equilibrium. We also compare the
results between a planet with a core and without a core.  To be
inflated to the same size, a planet
with a core, therefore possessing a larger gravitational binding
energy, needs a larger intrinsic luminosity $\cal L$ than a planet
with no core and the same mass.  We also show
that the opacity in the radiative envelope has a drastic effect on the
final equilibrium size of an inflated planet: the size would be much
smaller if grains are depleted in the radiative envelope.

We also consider the possibility of localized tidal dissipation.  Such
a process may occur in differentially rotating planets or near the
interface between the convective and radiative zones where the
wavelength associated with dynamical tidal response is comparable to
the density scale height. Localized dissipation may also occur through
the dissipation of resonant inertial waves or radiative damping in the
atmosphere.  In the strong shell-heating models, the one-to-one
relation between $K$ and $\cal L$ disappears because of the existence
of a radiative region caused by temperature inversion beneath the
shell-heating zone. The unheated planet's interior in such cases might
still be inflated due to a significant expansion of the gas above the
heating zone, although the overall expansion rate is less efficient
than that in the uniform heating cases as a result of a greater amount
of radiative loss from the planet's photosphere. Without gaining
entropy, the expanding interior cannot lift its degeneracy and
therefore cannot increase its elasticity. However, the gas above the
shell-heating zone can lift its non-ideal properties and hence enhance
its elasticity, leading to a decrease in $\Gamma$ in that region and
thereby diminishing the growth rate of $\cal L$ as the planet expands.

Finally, we consider the self consistent response, taking into account
the modification of heating rate due to a planet's expansion.  In this
paper, we adopt a constant-$Q$ prescription for equilibrium tides in
which the tidal dissipation rate is assumed to be proportional to
$R_p^5$.  The results for the uniform heating model suggest that a
young gaseous planet of $1~M_J$ without a solid core
can be thermally unstable and inflated
from $2~R_J$ to a size beyond $4~R_J$ if $e\sim 0.2$ at $a=10~\rsun$.
If the dissipation rate  is proportional to $R_p^5$,  and if most of the
tidal perturbation is deposited at $m/M_p=0.9$, a core-less young
planet of $1 M_J$ would be thermally unstable and inflated from $2 R_J$
to a size beyond $3 R_J$ for $e>0.07$ at $a=0.03$ AU or $e>0.26$ at
$a=10 \rsun$.

With the same heating concentration $m/M_p=0.9$ and $R_p$-dependence
in the dissipation rate, a young planet with a core at $a=0.03$ AU with 
an initial eccentricity $e>0.294$ can be inflated from $2~R_J$ to a
size beyond its Roche radius.  We have assumed that the convective
flow still behaves adiabatically even though the heating shell causes
a narrow radiative zone. However, the condition away from adiabaticity
implies that the internal heat is not transported away as
efficiently as in the case of  adiabatic convection, leading to a more
severe reduction in non-ideal properties of  the gas and therefore an even
faster decrease of $\gamma$ as the planet expands.  

Note that the Eddington approximation for the surface
boundary condition  is used in these models
rather than more detailed frequency-dependent model atmospheres.
This approximation is not necessarily valid for the
strongly irradiated atmospheres studied here (\cite{gs02}).
However it is unlikely to make much difference for
the main results discussed here, namely the behavior of the
planet's radius as a function of tidal dissipation energy.
It could, however, lead to errors in other kinds of predictions,
such as the radius as a function of opacity.

The equilibrium tidal dissipation formula is based on an {\it ad hoc}
assumption of a constant lag angle.  In reality,  the dynamical tidal
response of a planet through both gravity and inertial waves near the
planet's surface and convective envelope may be much more intense,
especially through global normal modes.  Their dissipation may provide
the dominant angular momentum transfer mechanism for the orbital
evolution and heating sources for the internal structure of close-in
extrasolar planets.  In the limit of small viscosity, the intensity
of tidal dissipation is highly frequency dependent (Ogilvie \& Lin
2004).  When the forcing and response frequencies are in resonance,
the energy dissipation rate is intense whereas between the resonances
it is negligible.  As the planets undergo structure adjustments, their
spin frequency, Brunt--V\"ais\"al\"a frequency distribution, the
adiabatic index, and equation of state also evolve.  Since all of
these physical effects contribute to the planets' dynamical response
to the tidal perturbation from their host stars, their response and
resonant frequencies are continually modified.  The results in this
paper indicate that the structure of the planet adjusts on a radiation
transfer time scale which generally differs from the time scale for a
planet to evolve through the non resonant region.  In addition, the
tidal forcing frequency also changes as the planets evolve toward a
state of synchronous spins and circular orbits.  Therefore, it is more
appropriate to consider a frequency averaged tidal dissipation rate.
In the limit of small viscosity, the frequency averaged dissipation
rate converges (Ogilvie \& Lin 2004) such that the equilibrium tidal
dissipation formula may be a reasonable approximation.  Nevertheless,
we cannot yet rule out the possibility that some close-in planets may
attain some non resonant configuration and stall their orbital
evolution.  Therefore, accurate measurement of the sizes of close-in
young Jupiters via planet transit surveys can be used to constrain the
theories of tidal dissipation and hence internal structure for these
objects.

\acknowledgements We wish to thank G. Laughlin, N. Murray, G. Ogilvie,
and E. Vishniac for useful conversation. We also thank the referee 
T. Guillot for the invaluable comments. Part of this work was
completed when one of us (PG) was a visitor at the UCO/Lick
Observatory, and he is grateful to K.-Y. Lo for the support of this
project. This work is supported by NSF and NASA through grants
AST-9987417 and NCC2-5418.

\appendix
\section{A Polytrope Model} 
In order to identify the dominant physical effects which determine the
internal structure of planets, we consider an inflated giant planet
that consists of a polytropic interior and a thin envelope. With this
model we can construct analytic solutions. For this analysis, we
assume that the polytropic interior comprises almost all of the mass
and radius, and that the thin envelope overlying the interior is
radiative and is composed of a non-degenerate ideal gas. Therefore,
the polytropic equation $P=K\rho^{1+1/n}$, together with the condition
for hydrostatic equilibrium, specifies the value of $K$ for a given
planet's mass $M_p$ and a given planet's size $R_p$ (\cite{cox}):
\begin{equation}
K =k_n M_p^{1-1/n} R_p^{-1+3/n}, \label{eq:K}
\end{equation}
where $k_n$ is a function of $n$.  All of thermodynamic quantities
must be continuous across the boundary between the convective interior
and the radiative envelope, such that
\begin{eqnarray}
&&P_b = {{\cal R}\rho_b T_b\over \mu} \label{eq:ideal}\\
&&P_b = K \rho_b^{1+1/n} \label{eq:polytrope}\\
&& \nabla_{polytrope}=\nabla_{rad} 
\approx {3\over 16\pi ac G}{P_b \kappa_b \over T_b^4}
{{\cal L} \over M_p}, \label{eq:radiative}
\end{eqnarray}
where the subscript $b$ denotes the values evaluated at the
boundary. The magnitude of $\nabla_{polytrope}$ is determined by the
equation of state of the gas in the interior such that it is a
function of $n$ rather than directly dependent on the magnitude of
$R_p$ and $M_p$. Note that in equation (\ref{eq:radiative}) we have
assumed that $\cal L$ does not vary greatly across the radiative
envelope after it emerges from the polytropic interior.  This
assumption should be a reasonable approximation for the thin radiative
envelope so long as there is no localized source of intense heating
there. The above four equations give rise to the relation
\begin{equation}
{\cal L} 
\propto {M_p T_b^3 \over \rho_b \kappa_b} \propto
{k_n^n M_p^n (T_b R_p)^{3-n} \over \kappa_b}
\left( {\mu \over {\cal R}} \right)^n, \label{eq:L}
\end{equation} 
where ${\cal R}$ is the gas constant and $\mu$ is the mean molecular weight.
From the opacity table, we find that the opacity at the boundary may
be roughly approximated by a power law
\begin{equation}
\kappa_b \propto {T_b}^a, \label{eq:kappa_b}
\end{equation}
with $a=4.08$.

\subsection{Completely Degenerate Interior}
The equation of state for a completely degenerate gas in the
non-relativistic regime is given by $P\propto \rho^{5/3}$, which
corresponds to the $n=3/2$ polytrope with a constant $K$. Therefore,
equation (\ref{eq:K}) leads to the well-known mass-radius relation for
low-mass white dwarfs:
\begin{equation}
R_p \propto M_p^{-1/3},
\end{equation}
which does not vary with ${\cal L}$. This independence is equivalent
to the expression
\begin{equation}
\gamma \equiv {d\ln {\cal L} \over d\ln R_p} \rightarrow \infty .
\label{eq:LR_deg}
\end{equation}
 
\subsection{Partially Degenerate Interior with $n=1$ Polytrope}
We now consider a model in which the planet  has a sufficiently
large $R_p$ such its interior is partially degenerate with an $n=1$
equation of state ($P \propto \rho^2$).  Partial degeneracy occurs
when $D\approx 1$. Therefore by setting $D=1$ in equation (\ref{eq:D}), 
the partially degenerate interior can be described by
\begin{equation}
\rho = c T^{3/2}, \label{eq:part_deg}
\end{equation}
where $c$ is a constant. Under these conditions, $\nabla_{ad} = 1/3$,
which is consistent with the numerical solution for the region
dominated by the pressure-ionized hydrogen atoms as shown in Figure
\ref{fig:nabla}.

At the boundary, equations (\ref{eq:K}), (\ref{eq:ideal}),
(\ref{eq:polytrope}), and (\ref{eq:part_deg}) uniquely determine the
value of $T_b$ for a given planet's size $R_p$ without having to consider 
the photosphere:
\begin{equation}
T_b =\left( {{\cal R} \over \mu} \right)^2 {1\over K^2 c^2} 
= \left( {{\cal R} \over \mu k_1} \right)^2 {1 \over R_p^4 c^2}
\approx 2300\,{\rm K} \left( {2R_J \over R_p} \right)^4 \left(
{2\times 10^{-7} {\rm g\,cm^{-3}\,K^{-3/2}}
\over c} \right)^2 \left( {1\over \mu} \right)^2.
\end{equation}
With equations (\ref{eq:L}) and (\ref{eq:kappa_b}), it follows that
\begin{equation}
{\cal L} \propto M_p R_p^{10.32} c^{4.16}. \label{eq:LR_partdeg}
\end{equation}

Figure \ref{fig:degeneracy} shows that the degree of degeneracy at the
center of the planet decreases from 21 to 9 as its radius expands from
1.777 $R_J$ to 2.826 $R_J$.  The value $c$ decreases as the degree of
partial degeneracy is reduced
($c \approx 6\times 10^{-9}$ g cm$^{-3}$ K$^{-3/2}$ when Fermi energy
equals thermal energy).  Therefore, in accordance with
equation (\ref{eq:part_deg}), the temperature of the planet can increase
with, even though $\rho$ decreases with, $R_p$.
\footnote{Extrapolating from equation (\ref{eq:part_deg}) to the case of high
degree of partial degeneracy, we find that, as energy is added to the
system, $\rho$ in equation (\ref{eq:part_deg}) remains essentially unchanged
while $T$ increases as $c$ decreases. The input energy is mostly
converted to an increase in temperature instead of doing the $PdV$
work in a degenerate state.} Temperature indeed increases until $R_p$
is increased up to $\approx 2.3 R_J$. The temperature then decreases
with $R_p$ after that, as seen in our numerical results. 

Figure
\ref{fig:delta} depicts the coefficient of thermal expansion at
constant pressure $\delta$ $(\equiv -(\partial \ln \rho /\partial \ln T)_P)$,
$\chi_T$ ($\equiv (\partial \ln P / \partial \ln T)_\rho$), and $J\delta$ 
(related to the adiabatic index, see eq(\ref{eq:entropy}))
for a
planet of $0.63M_J$ as a function of the co-moving mass coordinate $m/M_p$.
The boundaries $m/M_p=0$ and $m/M_p=1$ represent the location of the planet's
center and its photosphere, respectively.  
The curves in Figure \ref{fig:delta} are not smooth because the
calculation involves several numerical derivatives. The magnitude of
these thermal expansion coefficients decreases  as the  gas
changes its phase from the ideal to the non-ideal regime, and approach
to zero as a fluid increases its degeneracy, meaning that $PdV$ work
is of less importance in a more degenerate gas as indicated in
equation (\ref{eq:energy}).  Figure \ref{fig:delta} shows that $\delta$
and $\chi_T$ monotonically increase with $m/M_p$, resulting from a
transition from partial degeneracy in the inner region featuring the
pressure-ionized hydrogen gas, to the non-ideal phase of dense
molecular hydrogen in the middle range of $m/M_p$, and then to the
ideal-gas regime in the very outer region (not able to be shown
because of the small scale) where $\delta=\chi_T=1$.  The magnitudes of
$\delta$ and $\chi_T$ are larger in the case of a relatively large
size planet (with $R_p=3.22R_J$) because the equation of state for its
interior is less degenerate and more ideal than that for a smaller
planet (with $R_p=1.66R_J$). This correlation is consistent with the
consequence that the temperature variation results from the shift of
degeneracy. Since $c$ drops as $R_p$ increases, the term $c^{4.16}$ in
equation (\ref{eq:LR_partdeg}) indicates that the exponent of $R_p$ should
become less than 10.32 as the degree of partial degeneracy is reduced.

\subsection{Adiabatic interior composed of an ideal gas}
For planets with extremely large $R_p (> > 2 R_J)$, the density near
the center of the planet become sufficiently low that degeneracy is
lifted and the equation of state is better approximated by that of an
ideal gas for an $n=1.5$ polytrope.  In this limit, the radiative
envelope is also relatively extensive.  Integrating the radiative
diffusion equation over the radiative envelope, we find the ratio of
the temperature at the boundary $T_b$ to the temperature at the
photosphere $T_{ph}$ (\cite{cox}) to be
\begin{equation}
{T_b \over T_{ph}}=\left( {(1+n_{eff}) \nabla_{ph} -1 \over
(1+n_{eff}) \nabla_{ad} -1} \right)^{1\over m+s+4}, \label{eq:T_b/T_ph}
\end{equation}
where $\kappa \propto \rho^m T^{-s}$, $n_{eff}=(s+3)/(m+1)$,
and the value of $\nabla$ at the photosphere
\begin{equation}
\nabla_{ph} \equiv {3 \over 16\pi ac G}{P_{ph} \kappa_{ph} \over
T_{ph}^4}{{\cal L} \over M_p}
={1\over 8}{{\cal L} \over 4\pi R_p^2 \sigma T_{ph}^4 }.
\end{equation}
In deriving the above equation, we have used $P_{ph}=2GM_p/3R_p^2
\kappa_{ph}$.  Since ${\cal L}$ is usually much smaller than the
stellar irradiation $4\pi R_p^2 \sigma T_{ph}^4$ and therefore
$\nabla_{ph} <<1$, the ratio $T_b /T_{ph}$ is not a sensitive function
of ${\cal L}$ and $M_p$.  Since $T_{ph}$ is determined by the stellar
irradiation, neither it nor $T_b$ varies significantly with ${\cal L}$
and $M_p$.

There are some uncertainties for the values  of $m$ and $s$.  If we
evaluate $\kappa$ for the radiative/convective interface, we find from
equation (\ref{eq:kappa_b}) that $m=0$ and $s=-4.08$.  The temperature near
the surface layer is 2000-3000 K such that the diatomic molecular
hydrogen attains $n=5/2$ and $\nabla_{ad} =2/7$.  With these
parameters, we find from equation (\ref{eq:T_b/T_ph}), $T_b\approx 1.326
T_{ph} \approx 2150$ K when $T_{ph}=1620$ K. Near the photosphere, the
grains may also provide the dominant opacity source, in which case
$s=-1$ but other parameters have the same values. In this
case, we find from equation (\ref{eq:T_b/T_ph}) that $T_b\approx 1.91 T_{ph}
\approx 3090$ K which is too high for grain opacity to be relevant.
The actual values of $s$ and $T_b$ are probably between these extreme
cases.  Finally, with equations (\ref{eq:L}) and (\ref{eq:kappa_b}) we
find for $n=5/2$, an ideal-gas equation of state leads to 
\begin{equation} {\cal L} \propto M_p^{5/2}
R_p^{1/2}.\label{eq:LR_ideal}
\end{equation}

\subsection{Polytropic evolution and planet compressibility}
As we can see clearly from equations (\ref{eq:LR_deg}),
(\ref{eq:LR_partdeg}), and (\ref{eq:LR_ideal}), the values of $\gamma$
decrease from $\infty$ to  10.32 to 0.5 as the degeneracy is lifted for
increasing values of $R_p$. The physical properties responsible for
this change can be attributed to the ``compressibility'' of the
planet. It is well known that the compressibility of a gas is linked
to the effective exponent $\Gamma_{eff}\equiv d\ln P/d\ln \rho$ which
varies with different thermal conditions in a situation of thermal
equilibrium.  However, in the case of inflated giant planets,
$\Gamma_{eff}$ is always equal to the adiabatic value
$\Gamma_{ad}=(d\ln P/d\ln \rho)_{ad}$ due to an efficient energy
transport by convection.  Consequently, $\Gamma_{eff}$ varies with
different equations of state which are  characterized by the index of
polytrope $n$ in our case.  We shall elucidate this point in this
section.

In general, equations (\ref{eq:L}) and (\ref{eq:kappa_b}) can be
approximated with a simple relation
\begin{equation}
{\cal L} \propto {M_p \over \rho_b T_b}.
\end{equation}
The factor $M_p$ comes into the above expression because the larger
$M_p$ is, the larger gravity is in the thin radiative envelope, and
therefore the stronger $\cal L$ is due to the steeper temperature
gradient in the radiative envelope. Unlike the relation with $M_p$,
$\cal L$ decreases as $\rho_b$ increases simply because the radiative
energy flux decreases as the density and therefore the optical depth
increase.

The numerical solutions for the constant $\epsilon$ models show that
$T_b$ increases with $R_p$ when the degree of partial degeneracy is
relatively high.  But $T_b$ decreases with $R_p$ when the degree of
partial degeneracy is modest ($R_p$ is larger than 2.3 $R_J$).  This
tendency suggests that $T_b$, a thermal quantity introduced by the
equation of state in the framework of a polytropic analysis,
primarily varies with $c$ as shown in equation (\ref{eq:part_deg}).

\begin{figure}
\plottwo{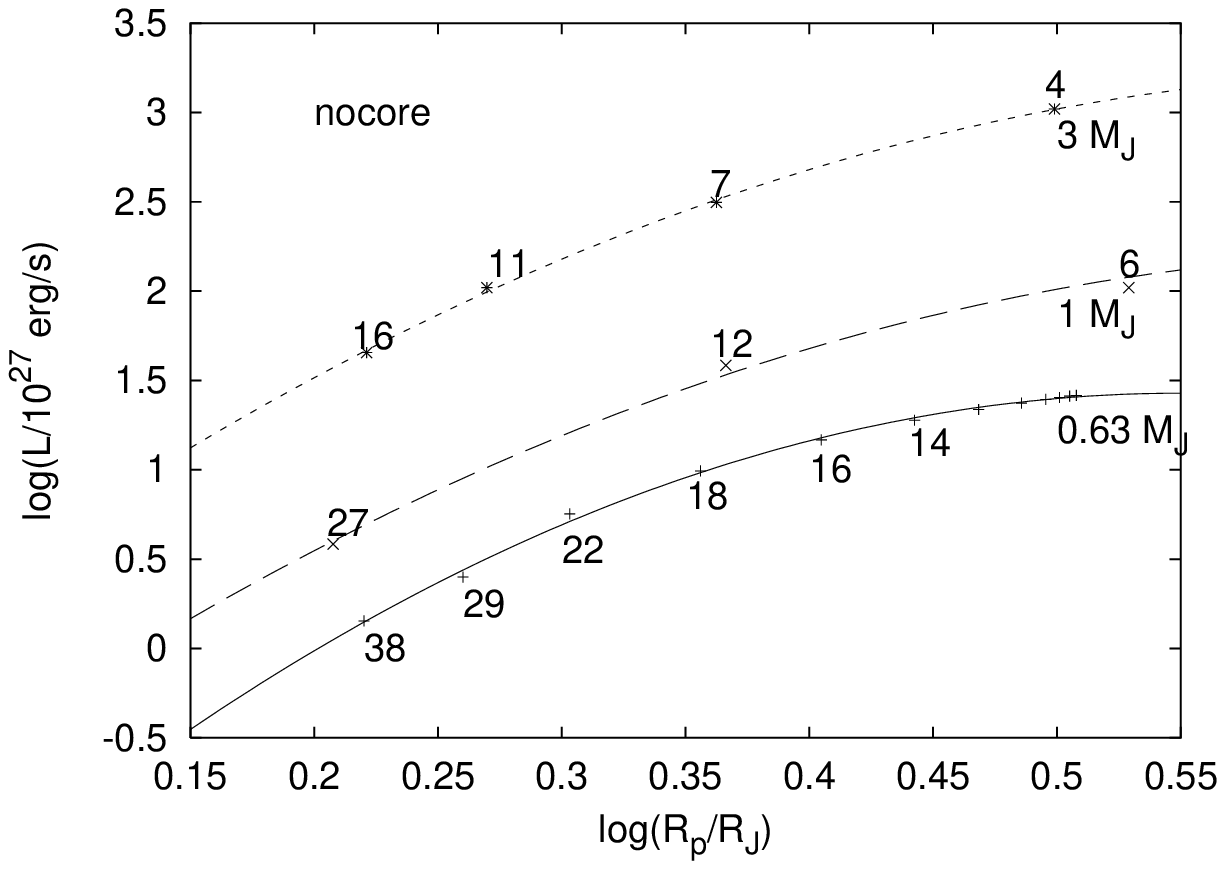}{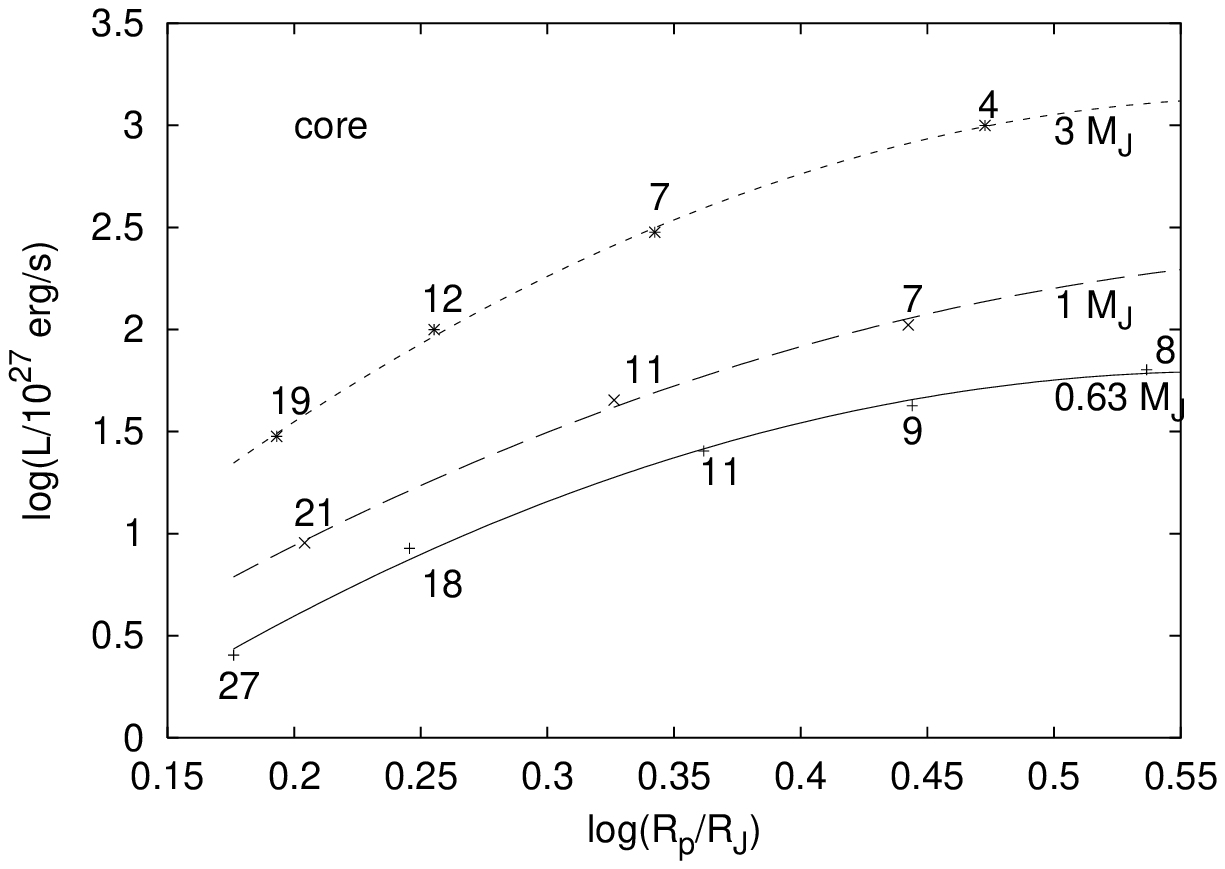}
\caption{Intrinsic luminosity as a function of planet's radius in
logarithmic scales for three different masses.  While the cases for
planets without a core are displaced in the left panel, the cases for
planets with a core are shown in the right panel.  Discrete points are
the simulation data, which are connected by fitting curves associated
with three different masses: $0.63M_J$ (solid line), $1M_J$ (dashed
line), and $3M_J$ (dotted line). The number marked next to each
discrete data point indicates the degeneracy $D$ at the planet's
center in the core-less cases (left panel) and the degeneracy $D$ on
the surface of the solid core in the core cases (right panel). All
three curves get flattened as the planet's size $R_p$ increases.}
\label{fig:luminosity}
\end{figure}

\begin{figure}
\plottwo{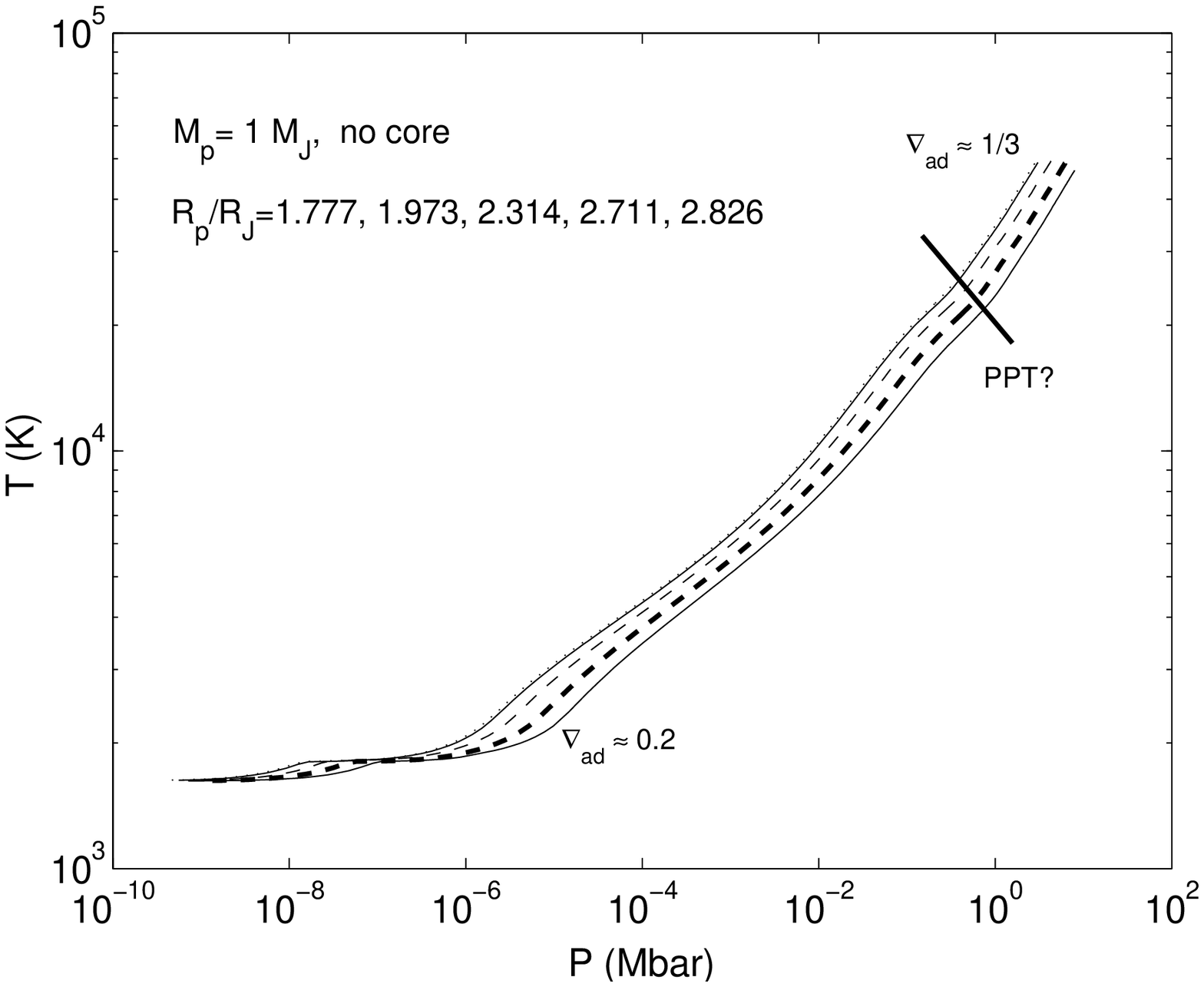}{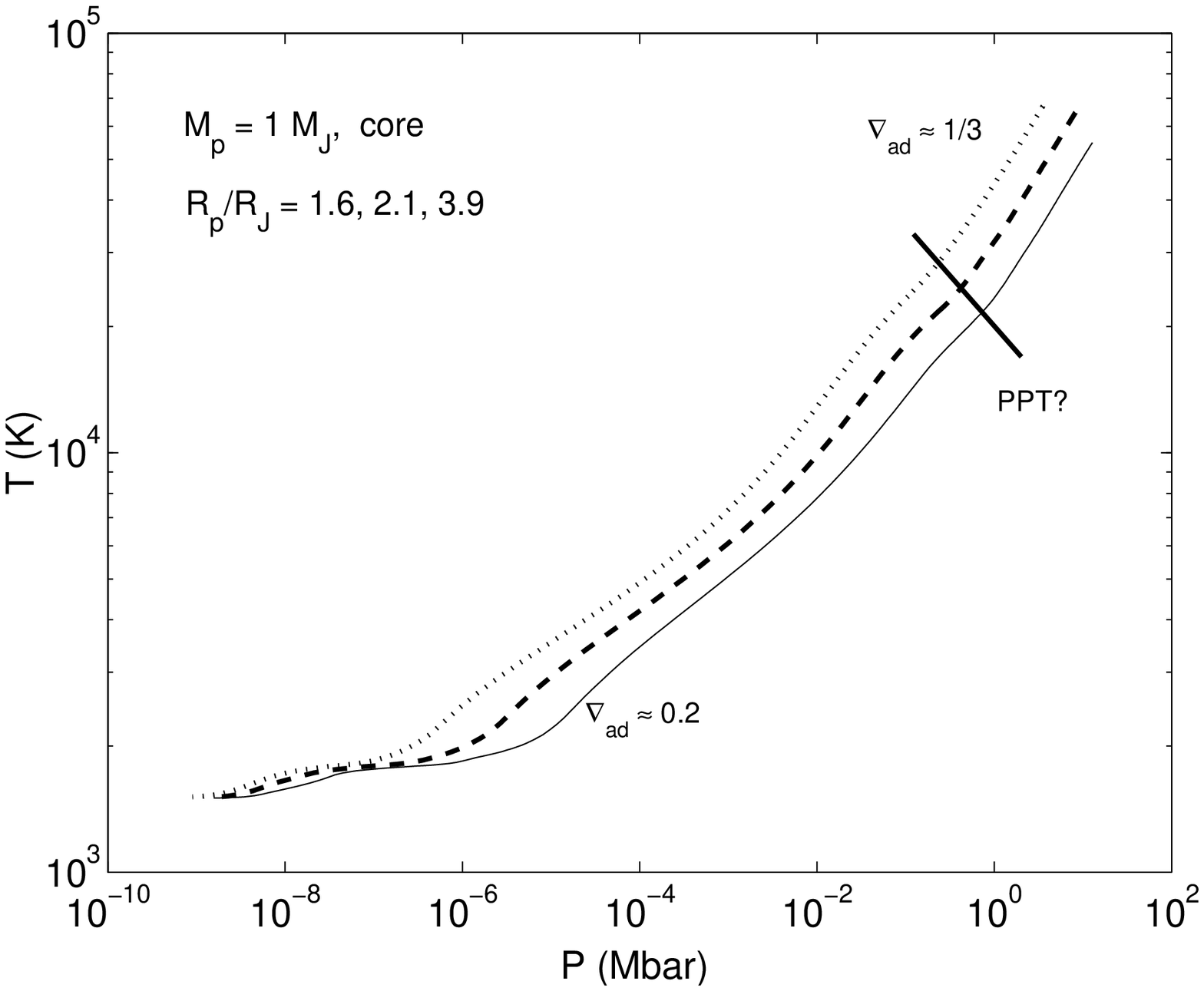}
\caption{The temperature profiles against the pressure inside a planet
of $1M_J$ with (right panel) and without (left panel) a
core. Different curves are plotted for different $R_p$: The interior
structure evolves from the lowest curve to the upper-most one as the
planet expands.}
\label{fig:nabla}
\end{figure}

Figure \ref{fig:rhob} illustrates the mass density at the
interface between the radiative envelope and the convective interior, 
$\log \rho_b$ as a function of the planet size $\log R_p$.  As the
planet expands, $\log \rho_b$ decreases but its slope gets
flattened. The similarity between Figure \ref{fig:rhob} and
\ref{fig:luminosity} suggests that the flattening of $\log \rho_b$ is
the major cause of  the flattening of $\log {\cal L}$ as the planet is
inflated. Equation(\ref{eq:L}) implies that
\begin{equation}
\rho_b \propto T_b^n M_p^{1-n} \left( {R_p^{n-3} \over k_n^n} \right)
\left( {{\cal R} \over \mu} \right)^n.
\label{eq:rho_b}
\end{equation}
As demonstrated in Figure \ref{fig:polytrope}, the simulation shows
that the fraction of the volume of the planet which deviates from the
$n=1$ polytrope increases with $R_p$.  This dependence is roughly
equivalent to the polytropic index being an increasing function of
$R_p$.  We fit the curves in the region of convection zones in Figure
\ref{fig:polytrope} for the core-less cases with a polytropic index and
show the results in Table \ref{tab:polytrope}.  The terms $({\cal
R}/\mu )^n$ and ${k_n}^n$ in the above equation increase with $n$, but
their influence is much weaker than the terms $R_p^{n-3}$ and
${M_p}^{1-n}$ because $R_p$ and $M_p$ are much larger. As a result,
the simple polytropic approach roughly suggests that ${\cal L} \propto
{R_p}^{3-n} {M_p}^n$, and hence the value of $\gamma$ (as defined
 by eq. \ref{eq:LR_deg}) decreases with $R_p$ as $3-n$ decreases with $R_p$.
Equation(\ref{eq:L}) gives a more precise representation of $\gamma$ as a
function of $R_p$:
\begin{eqnarray}
{d \gamma \over d \log R_p}
&=& 2{d n\over d \log R_p} \left( {d \log k_n \over d\log R_p}
-{d \log T_b \over d\log R_p} -1 \right)  \nonumber \\
&+& n{d^2 \log k_n \over d(\log R_p)^2}
-(a-3+n){d^2 \log T_b \over d(\log R_p)^2}
+{d^2 n \over d(\log R_p)^2}
\log \left( {M_p k_n \mu \over R_p {\cal R} T_b} \right) \nonumber \\
&\approx & -2{d n\over d \log R_p}.\label{eq:gamma_Rp}
\end{eqnarray}
The approximations made for the last expression in the above derivation
are that we ignore the terms associated with the 2nd derivative with
respect to $\log R_p$,  and that the terms $d\log k_n/ d\log R_p$ and
$d\log T_b/ d\log R_p$ are less than unity.

The magnitude of ${\cal L}$ increases with $M_p$ for a given $R_p$ as
suggested by the factor ${M_p}^n$, which is consistent with the
results shown in Figure \ref{fig:luminosity}.  An increase of $n$ with
$R_p$ in the polytropic analysis can be affirmed by the numerical
results which show that the central density $\rho_c$ and the central
pressure $P_c$ do not scale as fast as $R_p^{-3}$ and $R_p^{-4}$
respectively.  But, $\rho_c \propto R_p^{-1.6}$ and $P_c \propto
R_p^{-2.24}$, {\it i.e.}  the ratio of the central density to the
average density $\rho_c/<\rho>$ increases with $R_p$, meaning that $n$
also increases with $R_p$. Hence, the fact that an inflating planet
loses its partial degeneracy can be interpreted as an increase of $n$
from 1 in the polytropic analysis. The overall effect on the intrinsic
luminosity $\cal L$ for its dependence on $R_p$ is that $\gamma$
($\equiv d\ln {\cal L} /d \ln R_p$) decreases with $R_p$ due to the
flattening of the value of $\rho_b T_b$ as the planet inflates and 
loses its degeneracy.

Another piece of information which suggests a correlation between
electron degeneracy and $\gamma$ is the comparison of the evolution of
$\gamma$ between the core and core-less cases. As shown in \S3,
$\gamma$ passes 5 at a smaller $R_p$ for a less massive planet
($M_p=0.63$, $1M_J$) with a core than for that without a core. In the
case of a less massive planet, a planet with a core requires more
internal heating (see Fig.~\ref{fig:luminosity}) than that with no core
to be inflated to the same size, resulting in a lower degeneracy of
the planet interior and therefore a faster decrease in $\gamma$ even
though the core does not expand at all and the planet with 
a core is more gravitationally bound.  
On the other hand, $\gamma$
decreases below 5 when $R_p$ expands beyond $2.243R_J$ for a massive
planet ($M_p=3M_J$) in both core and core-less cases. The reason for
this independence of the core structure is because a $3M_J$ planet
without a core has a central density comparable to the 5 g/cm$^3$
which we impose for the density of the core in our numerical
prescription; i.e.  the interior structure of a massive core-less
planet is comparable to that of the planet with a core, resulting in
the similar $\cal L$ and $D$ required for the core and core-less cases
to be inflated to the same size  (see Fig.~\ref{fig:luminosity}).  In
summary, given a size of an inflated less massive
giant planet and assuming  that the
core is not heated and does not radiate, the planet with a
core is less degenerate than the one without a core,  
and hence the planet with a core has a smaller $\gamma$ and
expands more
for a given amount of entropy input (see equation (\ref{eq:C_M})
and detailed explanations below). 

If we trace back through the above derivation, we note that the relation
${\cal L} \propto {R_p}^{3-n} {M_p}^n$ originates from
equation (\ref{eq:K}). The parameter $K$ might be related to the total
entropy content of a polytropic object. The energy equation can be
expressed as follows (\cite{kw}):
\begin{equation}
ds=c_p J \left( d\ln P - {1\over J \delta} d\ln \rho
\right), \label{eq:entropy}
\end{equation}
where $J\equiv -\nabla_{ad}+1/\chi_T$ and
$\chi_T \equiv ( \partial \ln P / 
\partial \ln T)_{\rho}$. Therefore the adiabatic
exponent $\Gamma$ and the polytropic index
might be written in terms of $\delta$ and $\chi_T$:
\begin{equation}
1+{1\over n} =\Gamma ={1\over J\delta}. \label{eq:n_delta}
\end{equation}
For a completely degenerate gas, $\chi_T \rightarrow 0$, $\delta
\rightarrow 0$, and $1/J\delta$ can be written as the expression $d\ln
P /d\ln \rho$ which equals 5/3 in the non-relativistic case. As the
degree of degeneracy is reduced, both $\delta$ and $\chi_T$ increase
away from zero, and $J$, whose evolution is dominated by $1/\chi_T$,
decreases from infinity.  In the region of the middle and larger
values of $m/M_p$ where molecular hydrogen is abundant, both $\delta$ and
$\chi_T$ increase with $R_p$ as the non-ideal effect is lifted
(\cite{SCV})\footnote{The effect that $\delta$ increases as the
molecular gas reduces its non-ideal properties can be seen from Fig 8
in \cite{SCV} by noting that $\delta =\chi_T/\chi_{\rho}$, where
$\chi_{\rho} \equiv (\partial \ln P/\partial
\ln \rho )_T$. The rise of $\chi_T$ above 1 during the inflation
of a young giant planet is caused by the effects of molecular
dissociation (see Fig 7 in \cite{SCV} for more details).}.  This qualitative 
evolution of $\delta$ and $\chi_T$  is in agreement  with the 
plots obtained for a simulation with an inflated planet of 0.63
$M_p$ (first two panels  in Figure \ref{fig:delta}). The third panel in Figure
\ref{fig:delta} shows that the evolution of the product $J\delta$ is
dominated by $\delta$ and therefore increases as the planet expands,
leading to an increase in the  polytropic index $n$.

Motivated by the entropy equation (\ref{eq:entropy}), 
which indicates that the specific entropy $s\propto 
\ln (P/\rho^{\Gamma}) = \ln K$, 
we associate the issue of
entropy to the polytropic approach with a constant
$n$ by re-writing
equation (\ref{eq:K}) as
\begin{equation}
d \ln R_p = C_M d \ln K + C_K d \ln M_p, \label{eq:compress}
\end{equation}
where we have defined the compressibilities $C_M$
at constant mass
and $C_K$ at constant entropy as follows
\begin{eqnarray}
C_M &\equiv& \left( {d \ln R_p \over d\ln K} \right)_{M_p} = 
{n\over 3-n} \label{eq:C_M}, \\
C_K &\equiv& \left( {d \ln R_p \over d\ln M_p} \right)_K = 
{1-n \over 3-n}.
\end{eqnarray}
For a completely degenerate gas, $K=$constant (hence the result
$n/(3-n)$ shown in eq. (\ref{eq:C_M}) breaks down), $C_K=-1/3$ (since
$n=1.5$), and the term associated with $C_M$ vanishes in the
non-relativistic case. In the cases of terrestrial planets and
asteroids where the atomic/molecular interaction dominates and
therefore the density is on the similar order regardless of mass,
volume, and entropy, the interior structure can be roughly described
by the $n=0$ polytrope. When $n=0$, $C_K=1/3$, $C_M=0$ and
$\Gamma=\infty$.  These relations reflect a nearly constant density
($C_K=1/3$) for the planetary interior as well as the difficulty of
producing any non negligible density gradient with an entropy
($C_M=0$) and gravity ($\Gamma=\infty$) distribution.  In the special
case of partial degeneracy, $C_K \simeq 0$ and $C_M$ increases as $n$
rises from 1.\footnote{ Equation(\ref{eq:C_M}) gives the relation
$dC_M/d\log R_p = C_M^2 (3/n^2)dn/d \log R_p >0$. A more precise
expression derived from equation(\ref{eq:K}) is given by the equation
$dC_M/d\log R_p=C_M^2[(6/n^2)dn/d\log R_p - (2/n^3)(dn/d\log R_p)^2
\log (M_P/R_P^3) +(1/n^2)(d^2 n/d (\log R_p)^2) \log (M_P/R_P^3) +d^2
\log k_n/d (\log R_p)^2 ]$, which approximately equals
$C_M^2(6/n^2)dn/d\log R_p$ when the other terms associated with
$d^2/d(\log R_p)^2$ and $1/(\log R_p)^2$ are neglected.}  The zero
value of $C_K$ indicates a maximum size of the planet at $n=1$ solely
in response to the mass change (\cite{hubbard}).  Phenomenologically
this result arises from a transition from degeneracy ($M_p \propto
R^{-3}$) to the non-ideal equation of state ($M_p \propto R^3$) due to
atom-atom/molecule-molecule repulsion (\cite{bhll}; \cite{shu}).  
In terms of the polytropic index, the state of partial degeneracy
with $n\approx 1$ is just a transition from an $n=3/2$ degenerate
state to an $n=0$ constant-density state.  The rise of $C_M$ with $n$
means that, for a given mass, the rate of $R_p$ change in response to
the change of total entropy gets more sensitive as $n$ increases,
leading to the radius-adiabat relation and therefore the
radius-luminosity relation for constant heating per unit mass.

It is physically straightforward to see why $C_M$ increases with 
$n$ and (therefore decreases with $\Gamma$). It is because the adiabatic
exponent $\Gamma=\Gamma_{ad}$ is just
the bulk modulus for adiabatic expansion/compression.
Equation (\ref{eq:C_M}) can be re-arranged to have the
following form:
\begin{equation}
{1\over C_M}=3(\Gamma - \Gamma_{M_p}),
\end{equation}
where we have defined the bulk modulus at constant planet's mass
$\Gamma_{M_p} \equiv (\partial \ln P_c / \partial \ln \rho_c)_{M_p}$
which equals the constant 4/3.  The adiabatic bulk modulus
$\Gamma_{ad}$ should decrease as the planet's elasticity $C_M$
increases due to the reduction of electron degeneracy and non-ideal
effects. Figure~\ref{fig:Gamma_rho} illustrates the isothermal curves
for $\Gamma_{ad}$ as a function of $\log \rho$ in the case of hydrogen
(left panel) and helium (right panel).  The number marked on each
curve denotes the logarithmic value of temperature $\log T$(K). The
plots are drawn based on the tabulated equations of state
(\cite{SCV}).  The peaks of $\Gamma_{ad}$ ($\approx 2$) around $\rho
\gtrsim 1$ g/cm$^3$ at the temperature $T\sim 10^4$--$10^5$K roughly
correspond to the pressure-ionized regime of a Jupiter interior.
This effect may be compared 
to the sudden rise of $\Gamma_{ad}$ for $\log T$(K)$=3.22$ at
high densities as a result of non-ideal effects in the dense molecular
(for hydrogen) or atomic (for helium) regime (also see Figs. 8 \& 15
in \cite{SCV}).  The variation of $\chi_{\rho} \equiv \partial \log
P/\partial \log \rho |_T$ and of $\Gamma_{ad}$ are quite similar; the
rise of $\Gamma_{ad}$ in the pressure-ionized phase in the case of
higher temperatures should also result from the non-ideal effects
due to the interactions between densely-packed hydrogen and helium
atoms, increasing the rigidity of the fluid.  After all,
equation (\ref{eq:compress}) describes the simple fact that the size of
a planet is in general determined by its gravity ($M_p$), its entropy
content ($K$), and the elastic properties of the planet responding to
gravity (described by $C_K$) and to the entropy content (quantified by
$C_M$).

\clearpage

\begin{deluxetable}{crrrrr}
\footnotesize
\tablecaption{Polytropic Fitting for An Inflated Giant Planet
of 1$M_J$ Without A Core
\label{tab:polytrope}}
\tablewidth{0pt}
\tablehead{
\colhead{$R_p/R_J$} & \colhead{1.777} & \colhead{1.973} 
& \colhead{2.314} & \colhead{2.711} & \colhead{2.826}
}
\startdata
$n$  & 2.058 & 2.141 & 2.222 & 2.358 & 2.392 \\
$({k_n})^n$  & 0.098 &  0.102 & 0.107 & 0.118 & 0.122 \\
\enddata
\end{deluxetable}

\begin{deluxetable}{crrrrrrrrr}
\footnotesize
\tablecaption{Parameters for various models: $\epsilon_s=5\times
   10^{-6} \Lsun$, $M_p=1M_J$, and $T_0=1500$ K. The last column entitled ``figure''
   shows the figures, in terms of their labels, in which the model is plotted.
   For instance, Model 1 appears in Figures \ref{fig:rhoModels},
   \ref{fig:kappa}, and \ref{fig:Te}.
\label{tab:model}}
\tablewidth{0pt}
\tablehead{
\colhead{Model} & \colhead{$\epsilon$} & \colhead{${\epsilon_0
\over \epsilon_s}$} & \colhead{$\beta$} & \colhead{${m_0 \over M_p}$}
& \colhead{${\Delta m \over M_p}$}  & \colhead{${\kappa_d \over \kappa_0}$}
& \colhead{${T_e \over T_0}$} & \colhead{figure}
}
\startdata
1  & $\epsilon_1$ &  1     &  0 &   -    & -  &   1       &  1  & \ref{fig:rhoModels}, \ref{fig:kappa}, \ref{fig:Te}   \\
2  & $\epsilon_3$ &  1     &  - & 0.9999 & 0.00015 &   1       &  1  & \ref{fig:rhoModels}    \\
3  & $\epsilon_3$ &  1     &  - & 0.95   & 0.05 &   1       &  1  &  -  \\
4  & $\epsilon_3$ &  1     &  - & 0.90   & 0.05 &   1       &  1  &  \ref{fig:kappa}    \\
5  & $\epsilon_3$ &  1     &  - & 0.70   & 0.05 &   1       &  1  &  \ref{fig:rhoModels}   \\
6  & $\epsilon_3$ &  1     &  - & 0.05   & 0.05 &   1       &  1  &  \ref{fig:rhoModels}    \\
7  & $\epsilon_1$ &  1     &  2 &   -    &   -  &   1       &  1  & -    \\
8  & $\epsilon_2$ &  1     &  2 &   -    &   -  &   1       &  1  & -    \\
9  & $\epsilon_1$ & 10     &  0 &   -    &   -  &   1       &  1  & \ref{fig:case11+12}    \\
10 & $\epsilon_3$ & 10     &  - & 0.9999 & 0.05 &   1       &  1  &  \ref{fig:case11+12}    \\
11 & $\epsilon_3$ & 10     &  - & 0.90   & 0.05 &   1       &  1  &  \ref{fig:case11+12}   \\
12 & $\epsilon_3$ & 10     &  - & 0.70   & 0.05 &   1       &  1  &  \ref{fig:case11+12}, \ref{fig:Jdelta_12}    \\
13 & $\epsilon_1$ &  1     &  0 &   -    &   -  & $10^{-3}$ &  1  &  \ref{fig:kappa}    \\
14 & $\epsilon_3$ &  1     &  - & 0.90   & 0.05 & $10^{-3}$ &  1  &  \ref{fig:kappa}    \\
15 & $\epsilon_3$ &  1     &  - & 0.05   & 0.05 & $10^{-3}$ &  1  & -    \\
16 & $\epsilon_1$ &  1     &  0 &   -    &   -  &   1       & 0.8 &  \ref{fig:Te}     \\
17 & $\epsilon_1$ &  1     &  0 &   -    &   -  &   1       & 0.5 &  \ref{fig:Te}   \\
18 & $\epsilon_1$ &  $\propto R_p^5$     &  0 &   -    &   -  &   1       &  1  & - \\
19 & $\epsilon_3$ &  $\propto R_p^5$     &  - &  0.90  & 0.05 &   1       &  1  & - \\
20 & $\epsilon_3$ &  $\propto R_p^5$     &  - & 0.9999 & 0.05 &   1       &  1  & - \\

\enddata
\end{deluxetable}

\clearpage

\begin{figure}
\plottwo{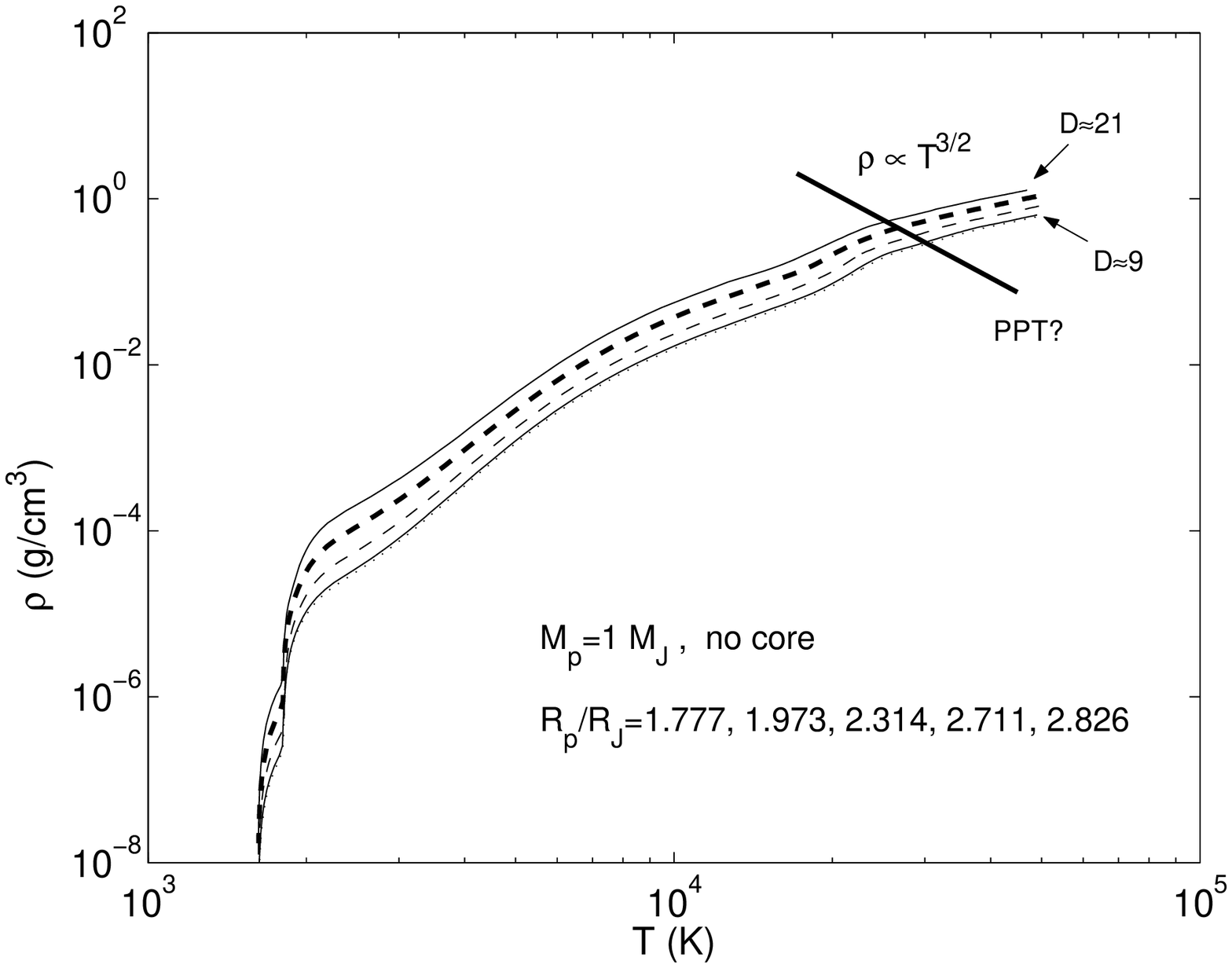}{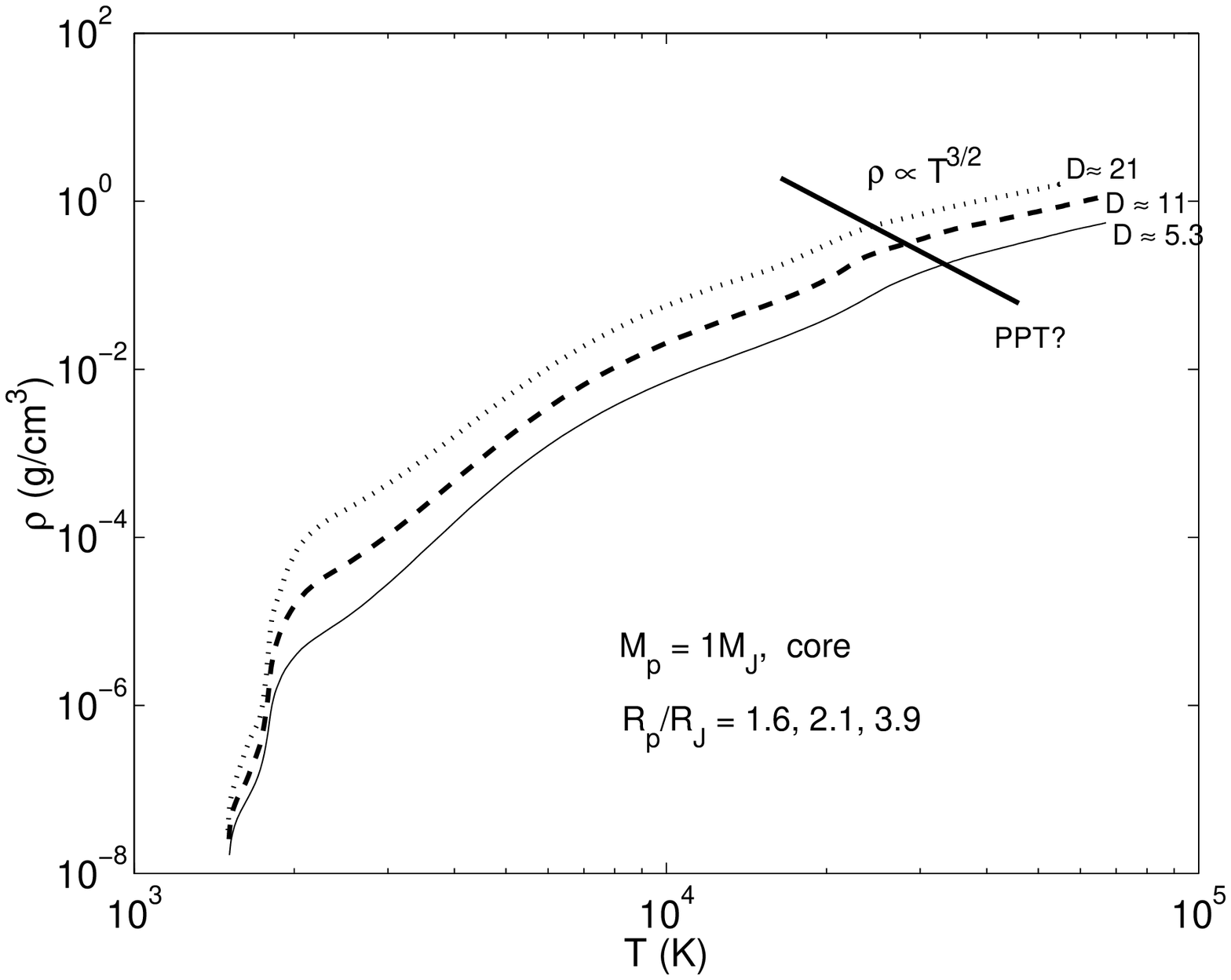}
\caption{ The mass density profiles as a function of the temperature
for various planet's sizes in the case of having a core (right panel)
and having no core (left panel).  The interior structure evolves from
the upper-most curve to the lowest one as the planet expands.  }
\label{fig:degeneracy}
\end{figure}

\begin{figure}
\plottwo{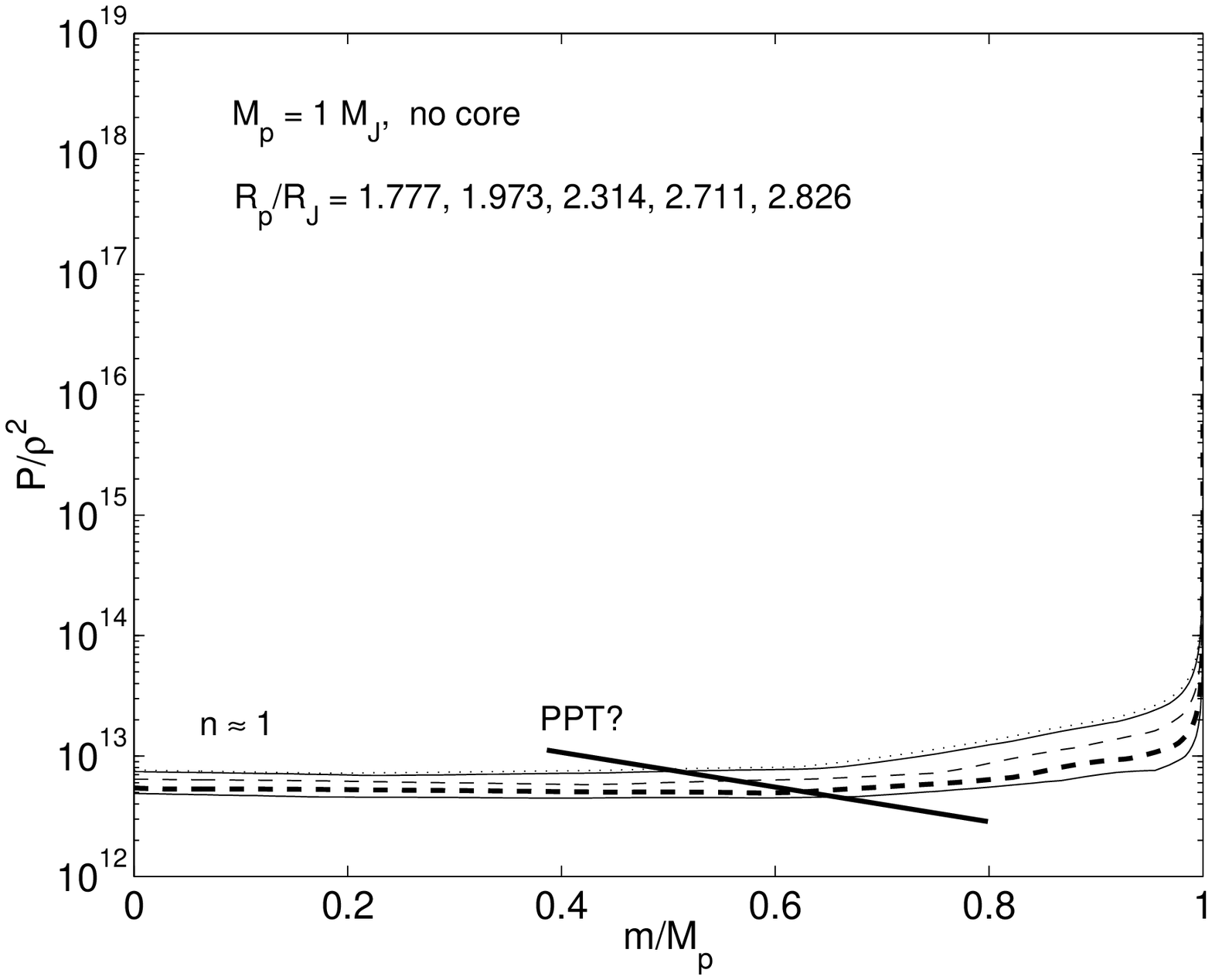}{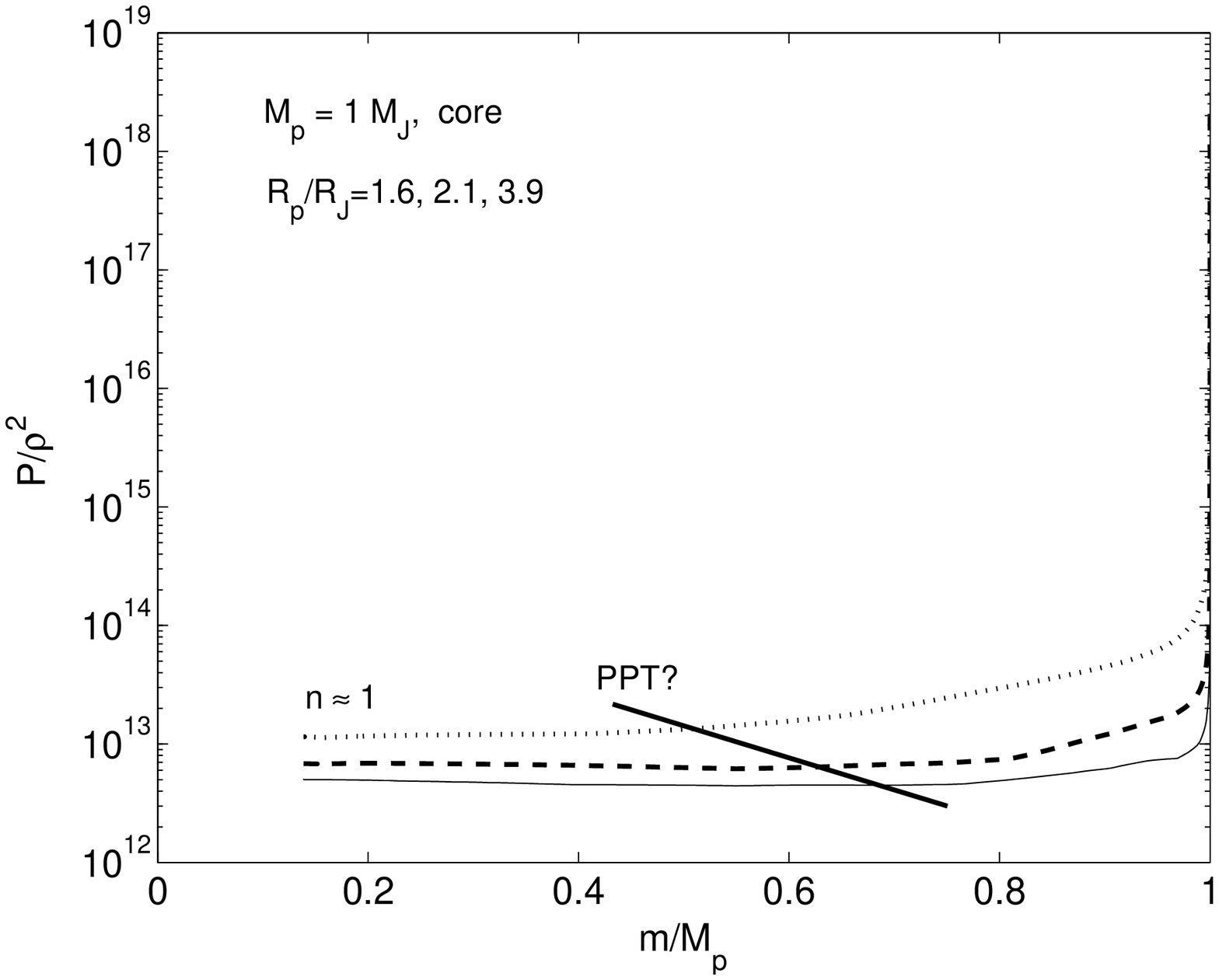}
\caption{ $P/\rho^2$ (in c.g.s units) as a function of the co-moving
radial coordinate $m/M_p$ for various planet's sizes in the cases of a
planet with a core (right panel) and without a core (left panel).
$m/M_p=0$ and $m/M_p=1$ represent the location of the planet's center and the
bottom of its photosphere respectively.  The interior structure
evolves from the lowest curve to the upper-most curve as the planet
expands to different sizes indicated by $R_p/R_J$ shown at the
upper-left corner of each panel.  }
\label{fig:polytrope}
\end{figure}

\begin{figure}
\plotone{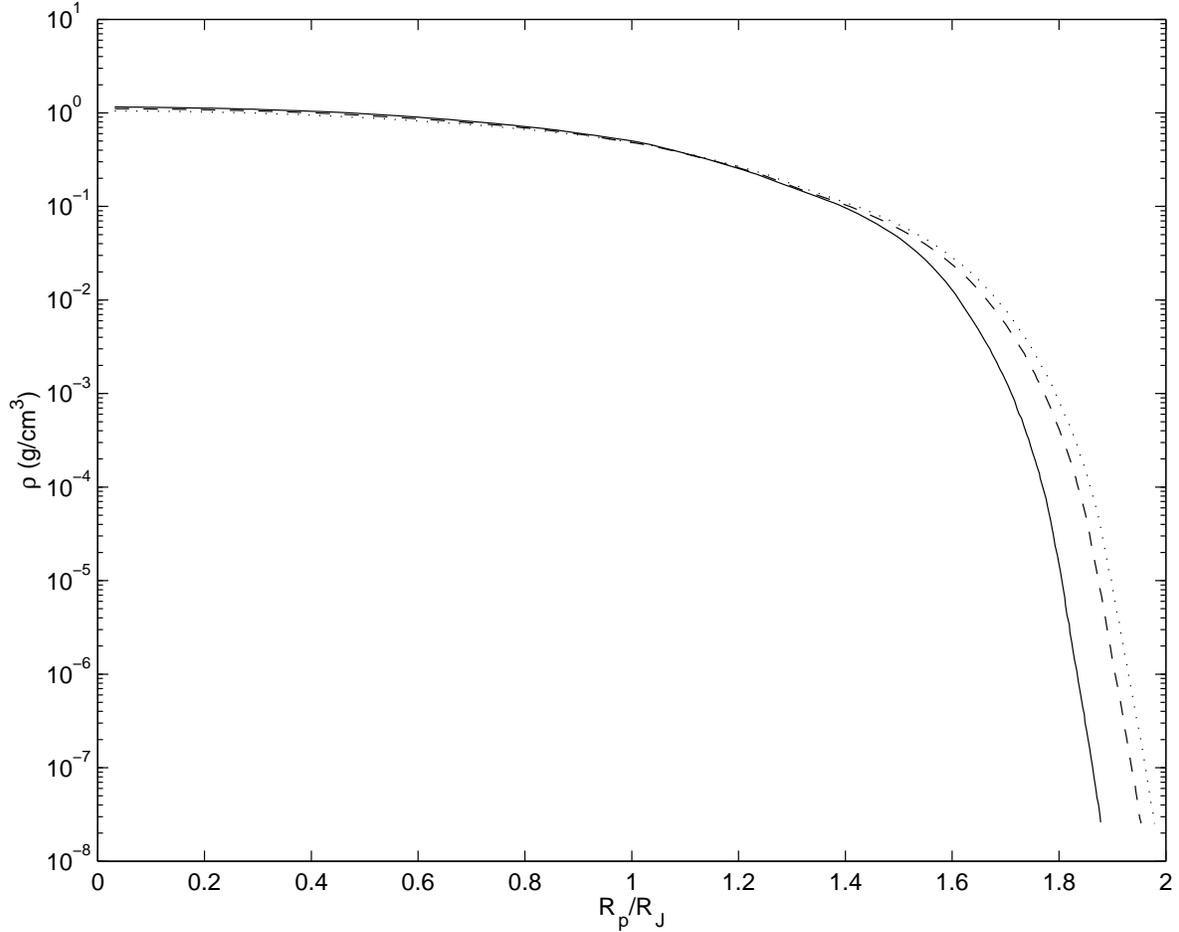}
\caption{The density profiles of a giant planet in the cases of Model
1 \& 6 (dotted curve), Model 2 (solid curve), and Model 5 (dashed
curve). Model 1 is a case of uniform heating,  while model 6 is
a case with the  same integrated heating rate but with
the energy deposition concentrated in a shell at mass
fraction $m_0/M_p=0.05$. Models 2 and 5 also have the same heating
rate but the energy is deposited in shells of mass
fraction $m_0/M_p=0.9999$ and 0.70, respectively.}
\label{fig:rhoModels}
\end{figure}

\begin{figure}
\plotone{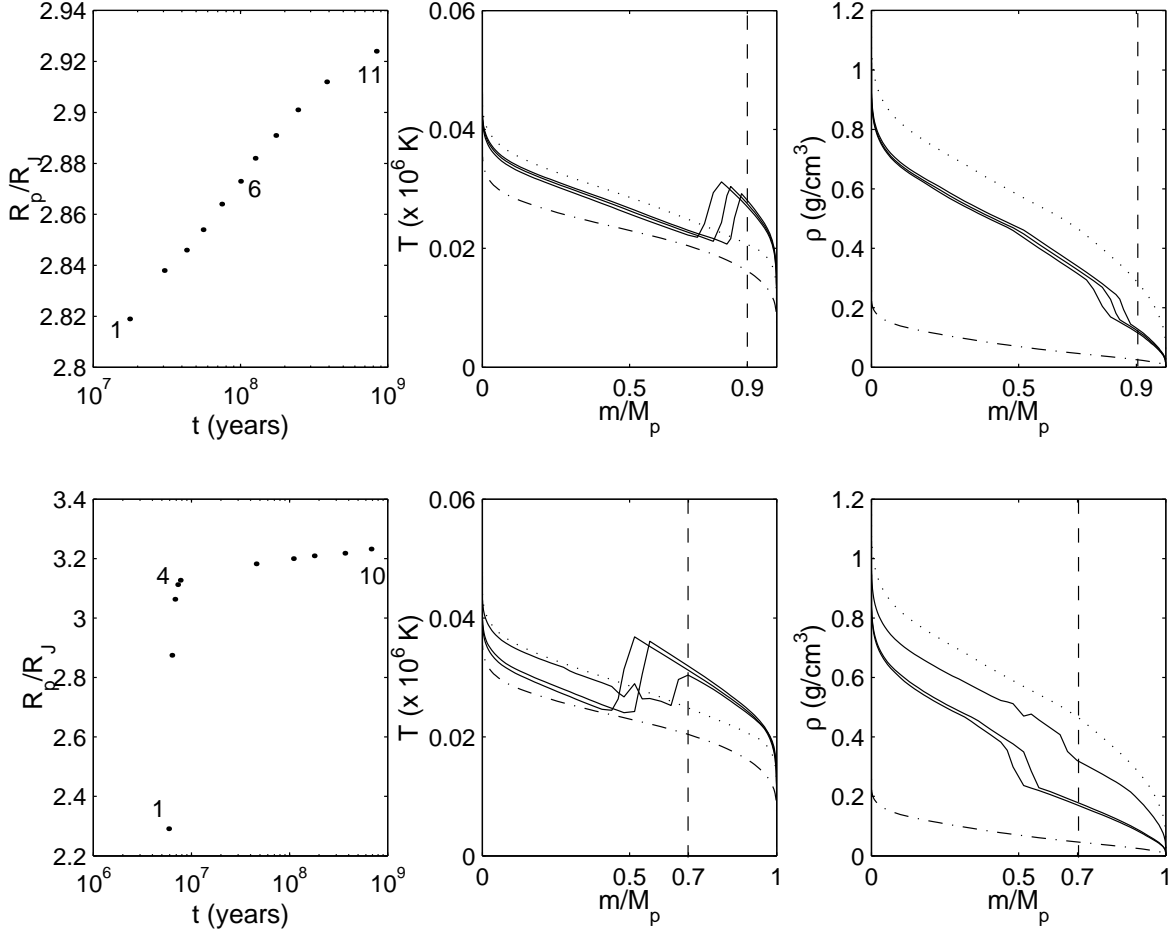}
\caption{The evolution of $R_p$, the temperature profile, and the
density profile for Models 11 (3 upper panels) and 12 (3 lower panels)
in Table \ref{tab:model}. While we only show the data after the planet
has reached a quasi-thermal equilibrium in Model 11, in Model 12 the
planet is not in the quasi-thermal equilibrium until the data point 4
(see the lower-left panel). The three solid curves from right to left
in each of the temperature and density plots for Model 11 (Model 12)
correspond to the three different stages marked by 1, 6, and 11 (1, 4,
and 10) respectively in the $R_p$ vs $t$ plot. The vertical dashed
lines in the temperature and density plots mark the location of the
maximal heating of the Gaussian heating profile for Model 11
($m_0/M_p=0.9$) and Model 12 ($m_0/M_p=0.7$). 
The temperature and
density profiles for the uniform heating per unit mass 
represented by Model 9 (dash-dotted line) 
and for the surface heating denoted by Model 10 (dotted
line) are also plotted for comparison.}
\label{fig:case11+12}
\end{figure}

\begin{figure}
\plotone{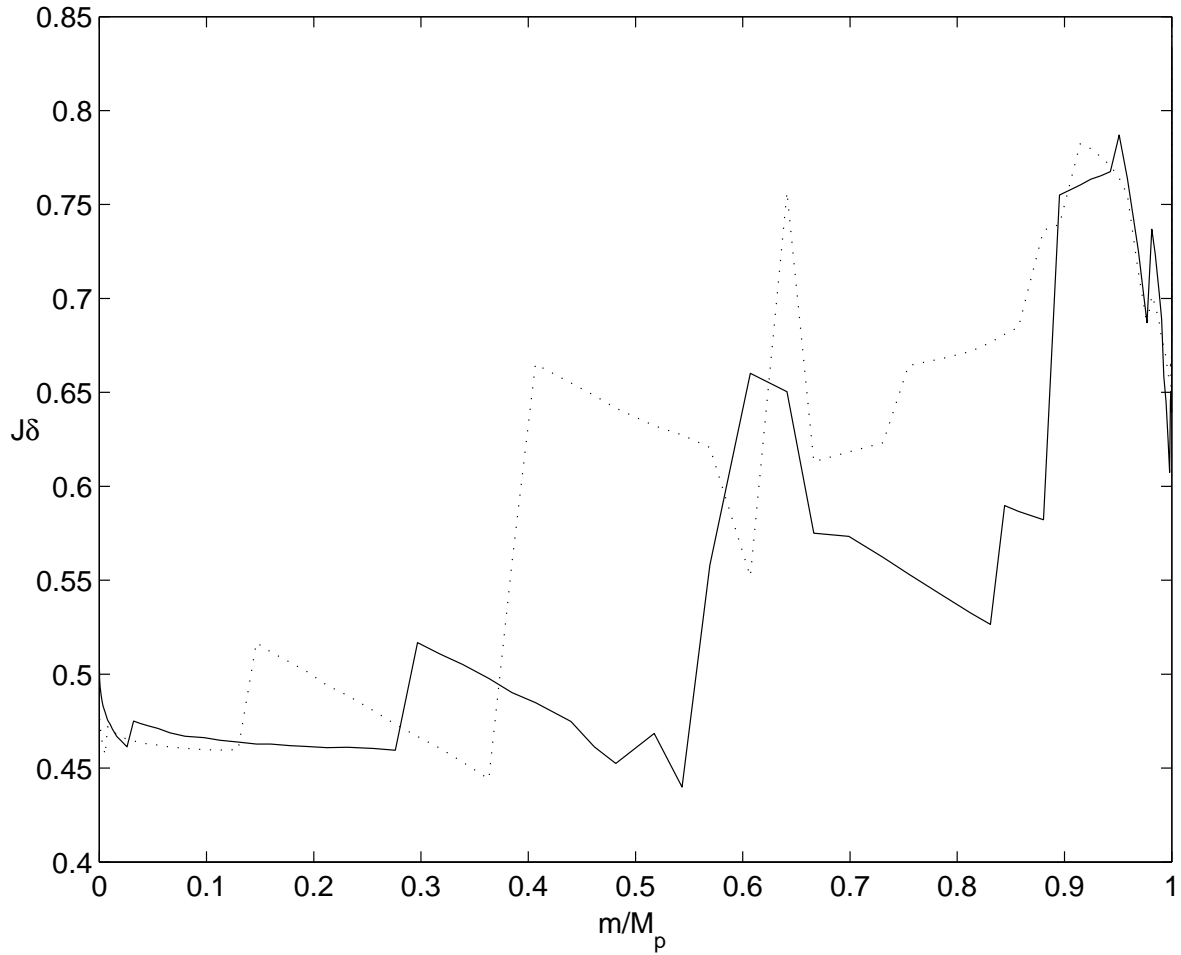}
\caption{The radial profiles of $J\delta$ ($=1/\Gamma$)
for the data point 1 (solid
curve) and the data point 4 (dotted curve) in Model 12. The data
points 1 and 4 are indicated on the lower-left panel
in Figure~\ref{fig:case11+12}.  In Model 12,
the energy is deposited in the shell of mass
fraction $m_0/M_p=0.7$.}
\label{fig:Jdelta_12}
\end{figure}

\begin{figure}
\plotone{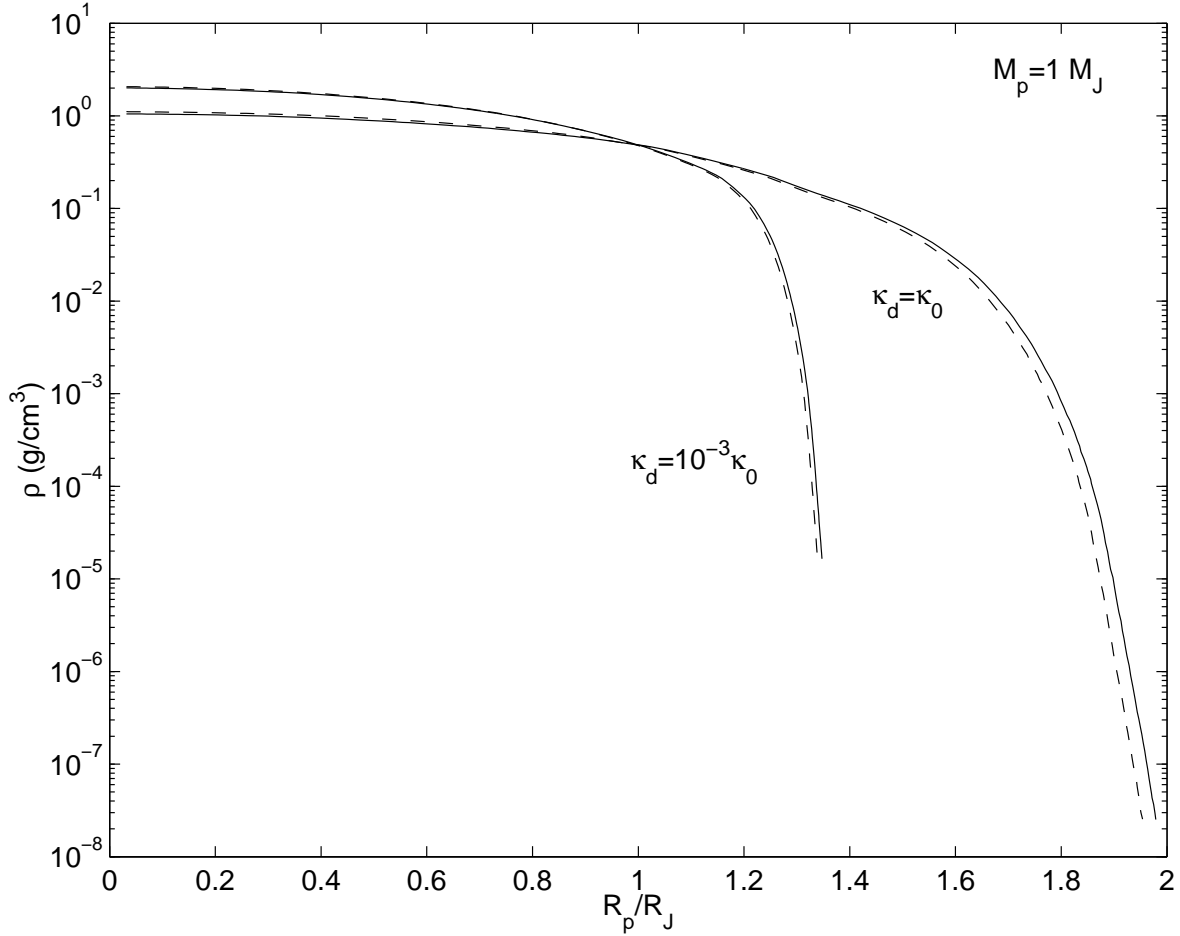}
\caption{The comparison of the density profiles for the regular grain
opacity $\kappa_d=\kappa_0$ (the solid curve for Model 1 and the
dashed curve for Model 4) and for the reduced grain opacity
$\kappa_d=10^{-3} \kappa_0$ (the solid curve for Model 13 and the
dashed curve for Model 14). While the heating is deposited uniformly in
mass in Models 1 and 13, Model 4 and 14 have
the same integrated heating rate but with
the energy deposition concentrated in a shell
at mass fraction $m_0/M_p=0.9$.}
\label{fig:kappa}
\end{figure}

\begin{figure}
\plotone{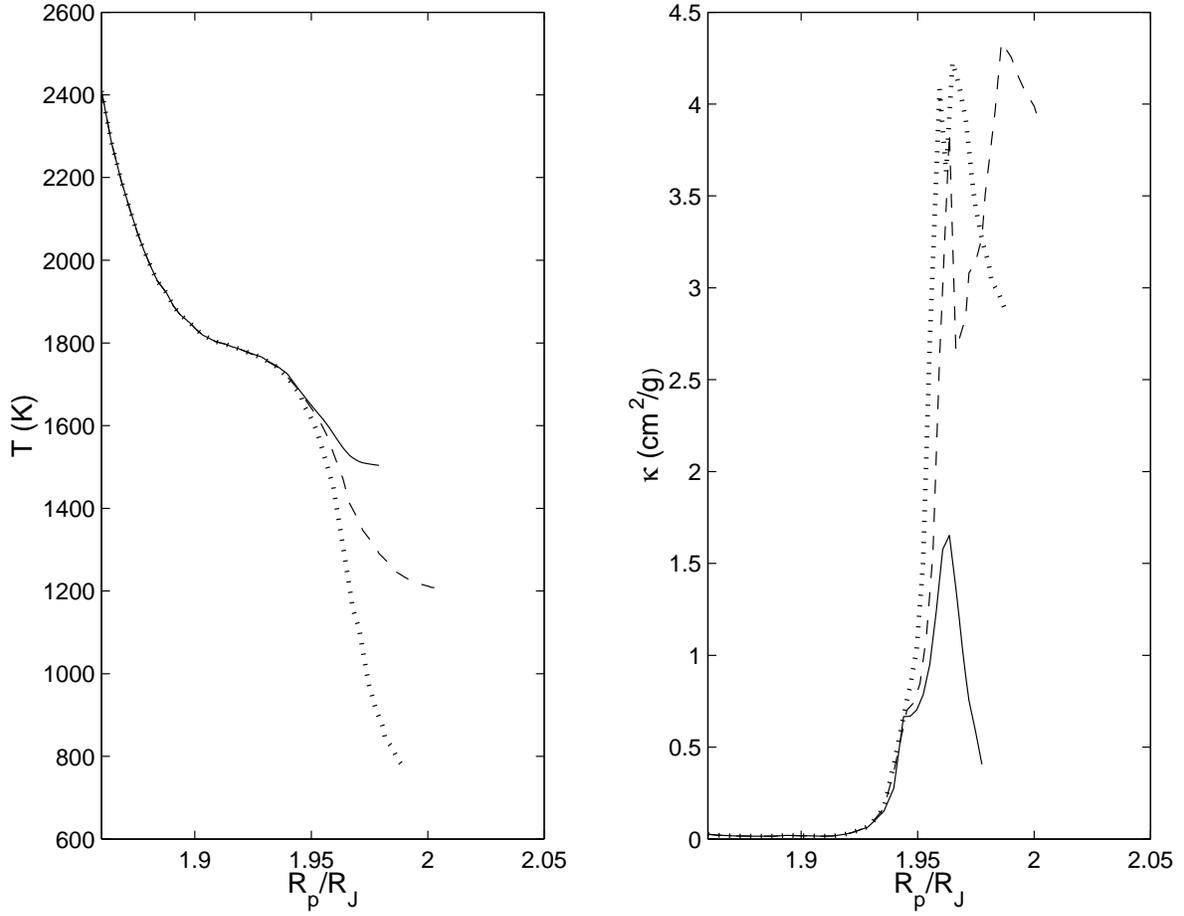}
\caption{Temperature and opacity profiles in the radiative envelope
for Model 1 (solid curve), Model 16 (dashed curve), and Model 17
(dotted curve). Different values of irradiation temperature
are imposed here: $T_0=1500$ K for Model 1, $0.8T_0$ for Model 16, and
$0.5T_0$ for Model 17.}
\label{fig:Te}
\end{figure}

\begin{figure}
\plotone{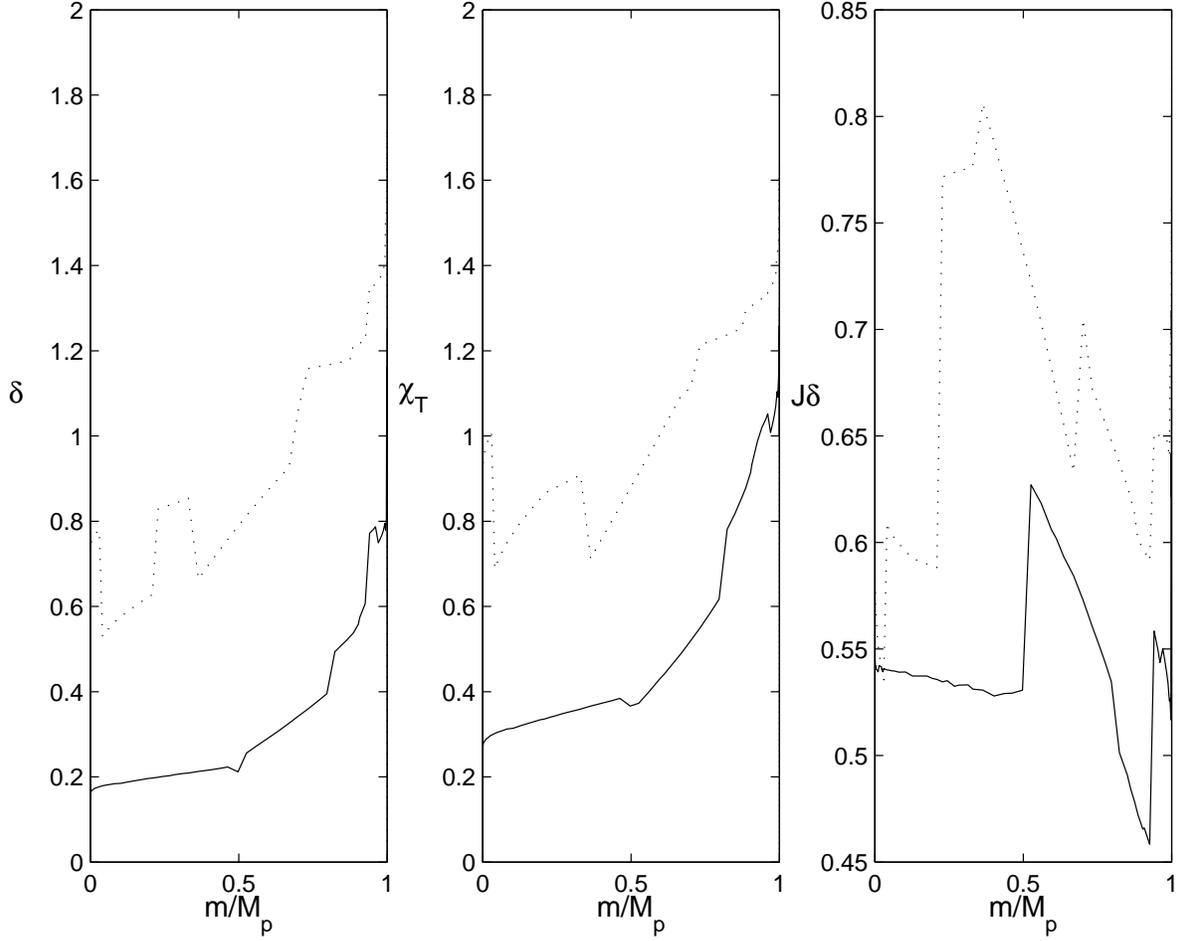}
\caption{The coefficient of thermal expansion $\delta$,
$\chi_T$, and $J\delta$ as a function of the co-moving mass coordinate $m/M_p$,
plotted here for an inflated planet of $0.63M_J$ with two different
radii: $1.66R_J$ (solid line) and $3.22R_J$ (dotted line).}
\label{fig:delta}
\end{figure}

\begin{figure}
\plotone{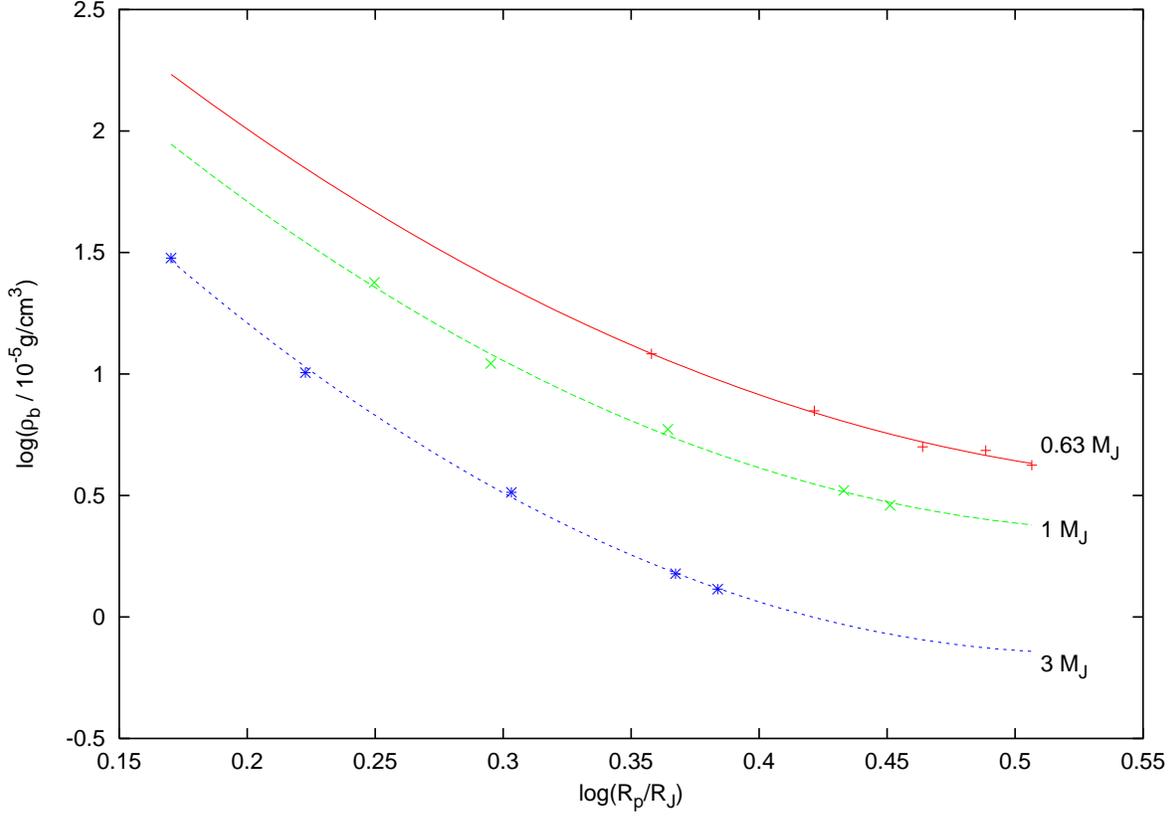}
\caption{The density at the interface between the radiative envelope
and the convective interior $\rho_b$ as a function of the planet's
radius $R_p$. While discrete points are the simulation data, the
fitting curves are illustrated by a solid line ($0.63M_J$), a dashed
line ($1M_J$), and a dotted line ($3M_J$).}
\label{fig:rhob}
\end{figure}

\begin{figure}
\plottwo{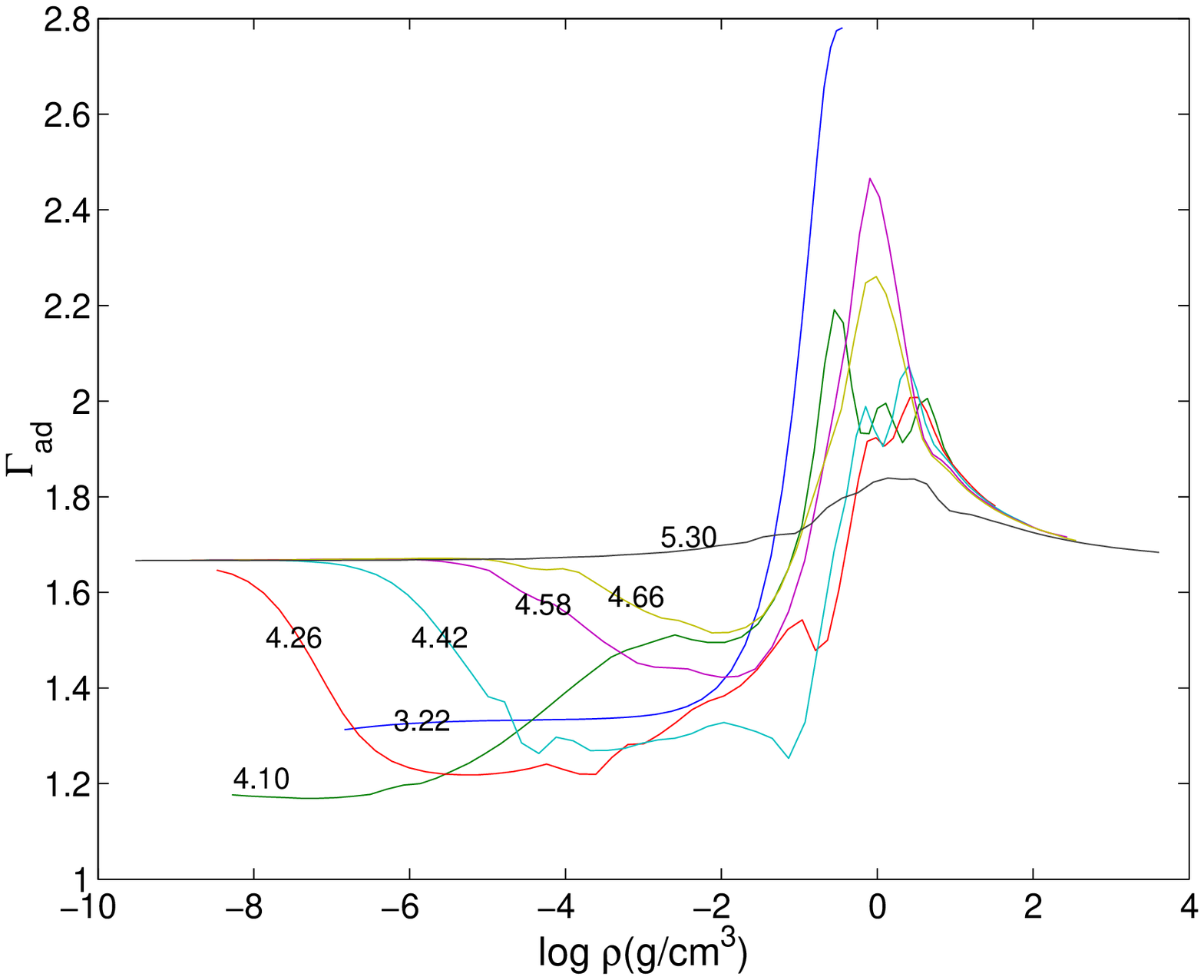}{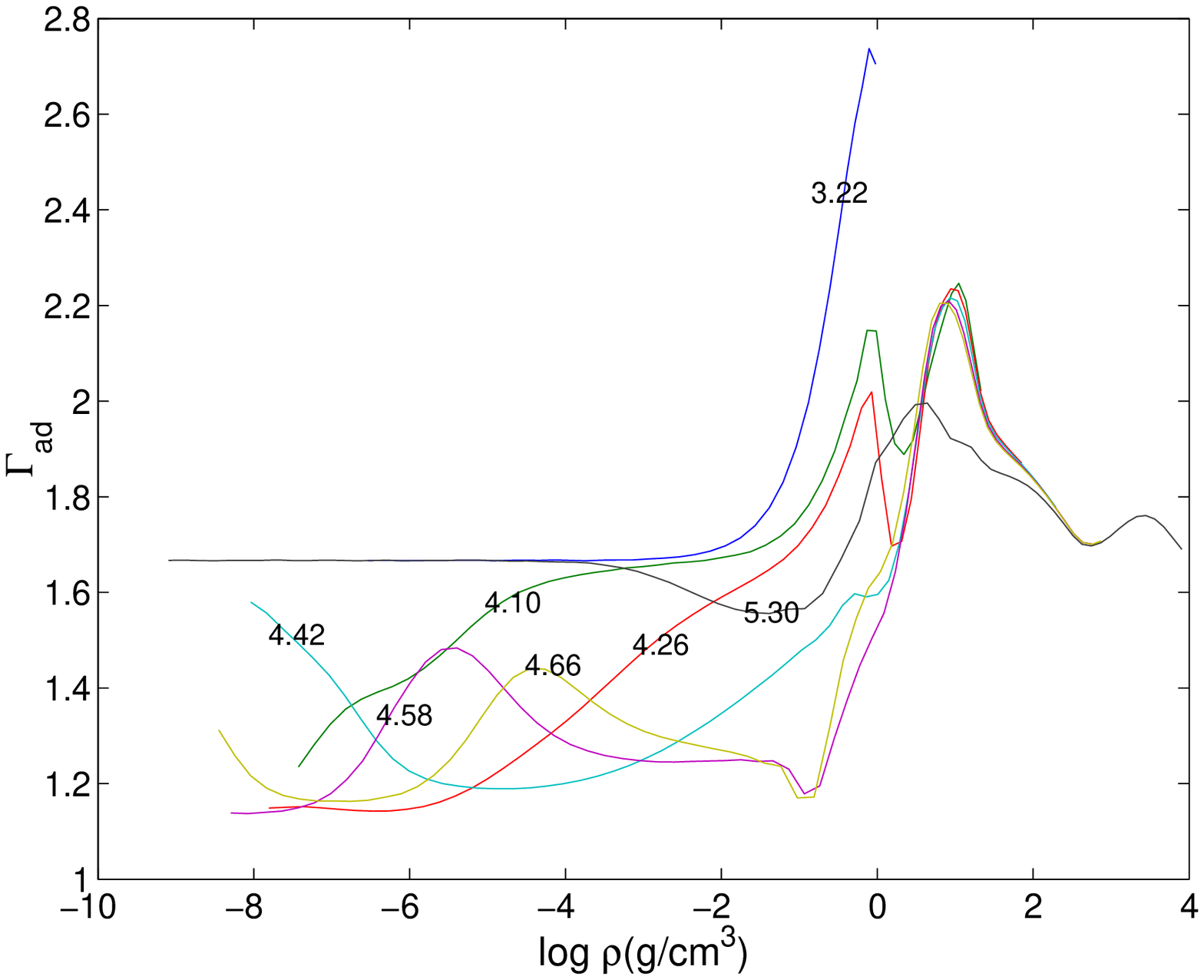}
\caption{$\Gamma_{ad}$ as a function of $\log \rho$(g/cm$^3$) and
$\log T$(K) for hydrogen (left panel) and for helium (right panel).
The number marked on each curve denotes $\log T$(K).}
\label{fig:Gamma_rho}
\end{figure}

\end{document}